\documentclass[draft,tightenlines,nofootinbib,preprint,aps,eqsecnum,amsmath,amssymb]{revtex4}

\newcommand{\beq}{\begin{equation}}
\newcommand{\eeq}{\end{equation}}
\newcommand{\bea}{\begin{eqnarray}}
\newcommand{\eea}{\end{eqnarray}}

\newcommand{\sgn}{\epsilon}

\begin{document}

\title{Dynamical Emergence of Instantaneous 3-Spaces in a Class of
Models of General Relativity.}

\medskip

\author{Luca Lusanna}

\affiliation{Sezione INFN di Firenze\\ Polo Scientifico\\ Via Sansone 1\\
50019 Sesto Fiorentino (FI), Italy\\ Phone: 0039-055-4572334\\
FAX: 0039-055-4572364\\ E-mail: lusanna@fi.infn.it}

\author{Massimo Pauri}

\affiliation{Dipartimento di Fisica\\
Universita' di Parma\\Parco Area  Scienze 7/A\\
 Sezione INFN di Milano Bicocca, Gruppo Collegato di Parma\\
 43100 Parma, Italy\\
 E-mail pauri@pr.infn.it}

\bigskip

\date{November 7, 2006, 12.30 am.}

\bigskip

\begin{abstract}

\noindent The Hamiltonian structure of General Relativity (GR), for
both metric and tetrad gravity in a definite continuous family of
space-times, is fully exploited in order to show that: i) the
\emph{Hole Argument} can be bypassed by means of a specific
\emph{physical individuation} of point-events of the space-time
manifold \emph{$M^4$ } in terms of the \emph{autonomous degrees of
freedom }of the vacuum gravitational field (\emph{Dirac
observables}), while the \emph{Leibniz equivalence} is reduced to
differences in the \emph{non-inertial appearances} (connected to
\emph{gauge} variables) of the same phenomena. ii) the
chrono-geometric structure of a solution of Einstein equations for
given, gauge-fixed, initial data (a \emph{3-geometry} satisfying the
relevant constraints on the Cauchy surface), can be interpreted as
an \emph{unfolding} in mathematical global time of a sequence of
\emph{achronal 3-spaces} characterized by \emph{dynamically
determined conventions} about distant simultaneity. This result
stands out as an important \emph{conceptual difference} with respect
to the standard chrono-geometrical view of Special Relativity (SR)
and allows, in a specific sense, for an \emph{endurantist}
interpretations of ordinary \emph{physical objects} in GR.

\bigskip
\bigskip

\noindent To appear in the book \emph{Relativity and the
Dimensionality of the World}, A. van der Merwe ed., Springer
Series \emph{Fundamental Theories of Physics}.

\end{abstract}

\maketitle

\vfill\eject

\section{Introduction}

The fact that, in the common sense view of ordinary life, phenomena
tend to be intuited and described in a non-relativistic
3-dimensional framework independent of any observer, is more or less
justified on the basis of the conjunction of the relative smallness
of ordinary velocities compared to the velocity of light and of the
neurophysiological capabilities of our brain concerning temporal
resolution.

Within the physical description of the world furnished by special
relativity theory (SR) in terms of the mathematical representation
of spatiotemporal phenomena in Minkowski space-time, one is
immediately confronted with the problem of defining the
3-dimensional instantaneous space in which ordinary phenomena should
be described.

As well known, the only possible resolution of this problem is based
on the \emph{relativization} of the description in relation to each
observer, as ideally represented by a time-like world-line. This
\emph{ideal} observer chooses an arbitrary \emph{convention} for the
synchronization of distant clocks, namely an arbitrary foliation of
space-time with space-like 3-surfaces: the instantaneous 3-spaces
identified by the convention. By exploiting any monotonically
increasing function of the world-line proper time and by defining
3-coordinates having their origin on the world-line on each
simultaneity 3-surfaces (i.e. a system of radar 4-coordinates), the
observer builds either an inertial or a non-inertial frame,
instantiating an observer- and frame-dependent notion of 3-space;
see Ref.\cite{1b} for the contemporary treatment of synchronization
and time comparisons in relativistic theories.

In the special case of inertial observers, the simplest way to
define distant simultaneity (of a given event with respect to the
observer) as well as to coordinatize space-time is to adopt the
so-called Einstein convention and exploit two-way light signals with
a single clock. Clearly, any two different observers must adopt the
same convention in order to describe phenomena in a coherent way
(essentially two different origins within the same frame).

The absolute chrono-geometrical structure of SR, together with the
existence of the conformal structure of light cones (Lorentz
signature) leads to the necessity of looking at Minkowski space-time
as a whole 4-dimensional unit.

\medskip

The way of dealing with such problems within general relativity (GR)
has been considered until now as embodying a further level of
complication because of the following facts:

i) The universal nature of gravitational interaction;

ii) The fact that the whole inertio-gravitational and
chrono-geometrical structure is jointly determined by the metric
field tensor \emph{g};

iii) The fact that, unlike SR, in GR we have a system of partial
differential equation for the dynamical determination of the
chrono-geometrical structure of space-time;

iv) The fact that the symmetry group of the theory is no longer a
Lie group like in SR but is the infinite group of diffeomorphisms in
a pseudo-Riemannian 4-dimensional differentiable manifold
\emph{$M^4$}. This fact, which expresses the \emph{general
covariance} of the theory, concomitantly gives rise to the so-called
\emph{Hole phenomenology} (from the famous \emph{Hole Argument},
formulated by Einstein in 1913, see Ref.\cite{a1}), which apparently
(in Einstein's words, see Ref.\cite{a2}) "Takes away from space and
time the last remnant of physical objectivity". Furthermore, it
renders the Einstein Lagrangian \emph{singular} with the consequence
that Einstein's equations do not constitute a hyperbolic system of
partial differential equations, a fact that makes the Cauchy initial
value problem almost intractable in the configuration space
\emph{$M^4$ }.

v) The absence of global inertial systems, which is a \emph{global
consequence} of the equivalence principle. Consequently, in
topologically trivial Einstein's space-times, the only globally
existing frames must be non-inertial.

vi) The necessity of selecting \emph{globally-hyperbolic}
space-times in order to get a notion of \emph{global mathematical
time} (to be replaced by a physical clock in experimental practice)
and avoid the so-called \emph{problem of time} \cite{35}.

\medskip

In this paper we will show that, at least for a definite continuous
family of models of GR - analyzed within the Hamiltonian framework -
it is possible to accomplish the following program:

\medskip

i) The metric field is naturally split into two distinct parts:

ia) An \emph{epistemic} part, corresponding to the arbitrary
constituents of the metric field (\emph{gauge} variables) that must
be completely fixed in order that the Hamilton-Einstein equations
became a well-defined hyperbolic system. This complete gauge-fixing
defines a \emph{global non-inertial spatiotemporal frame} (called
NIF) in which the true dynamics of the gravitational field must be
described with all of the \emph{generalized non-inertial effects}
made explicit. Once the NIF is fixed, the standard passive
4-diffeomorphisms are subdivided in two classes: those adapted to
the NIF, and those which are non-adapted: these latter modify only
the 4-coordinates in a way that is not adapted to the NIF.

ib) An \emph{ontic part}, corresponding to the \emph{autonomous
degrees of freedom} (2+2) of the vacuum gravitational field
(\emph{Dirac observables}, henceforth called DOs) expressed in that
NIF.

\medskip

ii) A \emph{physical individuation} of the point-events in
\emph{$M^4$} can be obtained in terms of the DOs in that NIF,  an
individuation that downgrades the philosophical bearing of the
\emph{Hole Argument} (see later). In Ref.\cite{27} we have shown
that matter does contribute indirectly to the procedure of physical
individuation, and we have suggested how this conceptual
individuation could in principle be implemented with a well-defined
empirical procedure, as a three-step experimental setup and protocol
for positioning and orientation.

\medskip

iii) A careful reading of the Hamiltonian framework leads to the
conclusion that the dynamical nature of the chrono-geometrical
structure of every Einstein space-time (or \emph{universe}) of the
family considered entails the existence of a \emph{dynamically
determined convention} about distant simultaneity. This is
tantamount to saying that every space-time of the family considered
is \emph{dynamically} generated in terms of a substructure of
embedded \emph{instantaneous 3-spaces} that foliates \emph{$M^4$ }
and defines an associated NIF (modulo gauge transformations, see
later). More precisely, once the Cauchy data, i.e. a
\emph{3-geometry} satisfying the relevant constraints, are assigned
in terms of the DOs on a \emph{initial Cauchy surface}, \emph{the
solution of the Einstein-Hamilton equations embodies the unfolding
in mathematical global time of a sequence of instantaneous 3-spaces
identified by the Cauchy data chosen}. As a matter of fact, such
3-spaces are obtained, through the solution of an inverse problem,
from the extrinsic curvature 3-tensor associated with the 4-metric
tensor, solution of the equations in the whole space-time
\emph{$M^4$}.

\bigskip

Consequently, in a given Einstein space-time, there is \emph{a
preferred dynamical convention} for clock synchronization that
should be used by every ideal observer (time-like world-line). In
practice, since the experimentalist is not aware of what Einstein
space-time he lives in, at the beginning the GR observers exploit
arbitrary conventions like in SR. Actually, it should be noted that
in GR, as in SR, the effective clock synchronization and the setting
of a 3-space grid of coordinates is realized by \emph{purely
chrono-geometrical means}\footnote{Note that in GR one cannot
ascertain in a simple way the usual relativistic effects concerning,
e.g., rods contraction and time dilation. In order to show such
effects one should have to exploit non-adapted coordinates
restricted to a small world-tube in which SR could be considered as
a good approximation.}. However, the preferred convention can be at
least locally identified by making a measurement of \emph{g} in the
4-coordinate system of the convention chosen and by solving the
inverse problem. In this way such GR observers could identify the
3-spaces and re-synchronize the clocks. Under the up-to-now
confirmed assumption that the real space-time is an Einstein space,
rather than a Weyl space, so that there is no \emph{second clock
effect} (see Ref.\cite{a3}), all of the clocks should maintain their
synchronization on every instantaneous 3-space. Of course, a true
physical coordinatization would require a dynamical treatment of
realistic matter clocks.

\bigskip

Finally, it must be stressed that, unlike the situation of temporal
ordering in SR, the unfolding of the 3-spaces constitutes here a
unique \emph{universal B-series ordering} of point-events\footnote{A
B-series of temporal determinations concerning events is
characterized by purely \emph{relational statements} like "earlier
than", "later than" and "simultaneous with". By contrast, an
A-series is characterized by \emph{monadic} attributes of single
events, like "future", "present" and "past". A-series sentences and
their truth-values depend upon the temporal perspective of the
utterer, while B-series sentences have truth-values that are
time-independent.}. Actually this holds true despite the fact that
the stratification of \emph{$M^4$ } in \emph{achronal 3-spaces } is
not gauge-independent. The point is that, \emph{on-shell}, every
dynamically admissible gauge transformation is the passive view of
an active diffeomorphism within a \emph{definite Einstein's
universe}: it changes the NIF, the Hamiltonian (with the
\emph{tidal} and \emph{inertial} effects), the world-lines of
material objects if present, and the same physical individuation of
point-events (see later), in such a way that the \emph{temporal
order} of \emph{any pair} of point-events and the
\emph{identification} of different material objects as to their
relative \emph{order} in space-time are not altered.

\medskip

Let us conclude with a few philosophical remarks or, rather,
specifications. In this paper we only deal with the {\it theoretical
properties} of space-time or space + time, {\it as mathematically
represented} in the three main space-time theories formulated in
modern physics, viz. the Newtonian, the special and the
general-relativistic. Accordingly, we do not take issue here with
such themes as temporal \emph{becoming} (absolute or relational),
tensed or de-tensed \emph{existence}, viewed as philosophical
questions \cite{a4} connected in particular to special or general
relativistic theories. Likewise, we exclude from the beginning any
statement concerning cosmological issues, either Newtonian or
general-relativistic. Above all, we are not concerned with the issue
of an alleged \emph{reification} of relativistic space-time as a
{\it real} four-dimensional continuum or a \emph{reification} of the
three-dimensional Newtonian space as the {\it Raum} of our
experience or phenomenal space in its various facets. We do believe
that such purposes are philosophically misled and grounded on an
untutored conflation of autonomous metaphysical issues with a
literal interpretation of physical theories.

Certainly in Minkowski space-time there is no absolute fact of the
matter as to which an event is \emph{present}. Yet there is no
absolute fact of the matter about presentness of events in Newtonian
absolute time either, as there is no fact of the matter about
presentness in \emph{any} physical \emph{theory} (though of course
not in the physicists' practice !). Being valid by assumption at any
time\footnote{Things seem a bit more complicated in cosmology, yet
not substantially.}, a physical theory cannot have the capacity of
singling out a particular moment as "the present". Likewise no
\emph{physical experiment} can be devised having the capacity of
telling whether a particular time signed by the hand of a clock is
"the present" or not. This kind of knowledge requires the
\emph{conscious awareness} of a living subject, see Ref.\cite{a5}.
However, no living observer can be forced within the Minkowski
space-time (or any general-relativistic space-time model). No living
observer can be there to collect in a \emph{factually possible}
(specifically \emph{causal}) way the infinite amount of information
spread on the space-like Cauchy surface which is necessary to solve
the initial value problem according to well-known mathematical
theorems and thereby defining the attributes of physical events.
There is no living observer who can act to selectively generate a
situation in their environment so that this situation, as a cause,
will, according to their causal knowledge, give rise later with
great probability to the effect which is desirable to them. Finally,
there can be no living observer with the freedom to check the very
empirical truth of the physical theory itself. Any alleged
reification leads to a notion of the world which includes
everything, in particular the object of an action and the agent of
the same action. \emph{This world, however, is a non-factual world
that nobody can "observe", study or control}.

\bigskip

These limitations, which are intrinsic to the nature of the
scientific image, can be easily misunderstood. For example, the fact
that Minkowski space-time must be considered as a whole
4-dimensional unit gave origin to the \emph{prima facie} appealing
but misleading notion of \emph{block} universe, which seemed to
entail that in Minkowski space-time there be no \emph{temporal
change}, on the grounds that there cannot be any \emph{motion in
time}. There is certainly no motion in time in the \emph{ordinary
sense of the term}, but \emph{change} is there even if necessarily
described in terms of the tenseless language of a
(observer-dependent) \emph{B-series}.

\medskip

The issue of motion in our context can be seriously misinterpreted.
The \emph{ordinary sense} of the term - the one used in particular
by the experimentalists in their laboratory - refers to the
observation that there is some object "moving". Since this "moving"
always takes place in "the present" of the experimentalist, as all
other phenomena "taking place", one could be easily misled to
believe that the \emph{physical essence} of the concept of motion
could be captured by the \emph{experience} of an object as "moving",
more or less on the same footing in which it is often asserted that
there is a "moving now". Of course, there are moments of the motion
history of the object that are \emph{now} "past" and characterized
as moments of a "remembered present". However, if the notion is no
longer taking place "now", we can only describe the former motion as
a purely B-series sequence of positions of the object at different
times (of a clock), a description which expresses exactly the only
\emph{objective physical meaning} of "motion". Accordingly, since
there is \emph{no transiency} of the "now" in Minkowski space-time,
only \emph{tenseless senses} of the words "becoming", "now" and even
of "motion" are admitted in the relativistic idiom. Certainly, the
ordinary sense of "motion" is perfectly legitimate in the
\emph{practice} of physics and, therefore, in the idiom of the
experimentalist. This entails that \emph{locally} and for a
\emph{limited length of time}, the experimentalist can project - so
to speak - his practical view of the motion into Minkowski
space-time as a mental aid for his intuition of the physical
process. The experimentalist can consider, e.g., a world-line of a
point-mass endowed with a clock and, as time goes by, sign on the
world-line certain times indicated by his own synchronized physical
clock in the laboratory, as a running chart. As long as the motion
takes place, this can be a useful way of reasoning. It should be
clear, however, that as soon as the limited allotted time has
expired, the intuitive spatiotemporal representation of the motion
given by the experimentalist is concluded and \emph{nothing remains}
in the Minkowski picture which is different by a section of a
standard infinite world-line with a B-series finite sequence of
marked events.
\medskip

On the other hand, this being said, could we suggest a specific case
in connection with the so-called \emph{endurantist/perdurantist}
debate ? Briefly, the contrast can be roughly summarized as follows:
the \emph{endurantist} takes objects (including people) as lacking
temporal extent and persisting by being \emph{wholly present} at
each moment of their history, while the \emph{perdurantist} takes
objects as persisting by being temporally extended and made up of
different \emph{temporal parts} at different times \cite{a6}. This
opposition, which could appear \emph{prima facie} as a mere matter
of terminology at the level of the scientific image, is
\emph{formally} vindicated within the orthodox view of relativistic
theories. Actually, if material objects were in any relevant sense
3-dimensional and persisted by occupying temporally unextended
spatiotemporal regions, how could they fit in with the unavoidable
4-dimensionality of relativistic space-time ?
\medskip

We believe, in general, that it is remarkably difficult, if not
structurally unsound, to devise any conclusive argument from physics
to metaphysical issues. It is true that four-dimensionalism, as a
philosophical stance, is sometimes used as a shield for perdurantism
and three--dimensionalism as a synonym for endurantism. Since,
however, we are avowedly averse from any kind of \emph{reification}
of the relativistic models of space-time, we should specify the
meaning, if any, of notions like \emph{endurantism} or
\emph{perdurantism} once restricted to the idiom we deem admissible
within the spatiotemporal scientific image. It is clear that,
whatever meaning we are ready to allow for the notion of
\emph{object} as representable in Minkowski space-time, it cannot be
\emph{wholly present} in any sense. Things, however, are apparently
different in GR, just in view of our results.

In our description of GR, the 4-dimensional spatiotemporal manifold
is \emph{dynamically foliated by gravitation} into global
\emph{achronal} 3-spaces at different global times. Therefore the
notion of a \emph{wholly present material object} becomes
\emph{compatible} with an \emph{endurantist} interpretation of
temporal identity. Note that we are avowedly using the restricted
adjective \emph{compatible}, because we do not want to be committed
to any specific philosophical stance about the issue of the
\emph{identity} of objects in general. We shall define a material
object (say a \emph{dust} filled sphere) as \emph{wholly present} at
a certain global time $\tau$ if all of the physical attributes of
its constituent events can be obtained by physical information
wholly contained in the structure of the 3-space at time $\tau$.
Note that such information contains in particular the 3-geometry of
the leaf and all the relevant properties of the matter distribution
that are also necessary for the formulation of the Cauchy problem of
the theory. Having adopted this definition, we will return to the
issue of the \emph{endurantism/perdurantism} dispute, restricted to
our formulation of GR, only at the end of the paper after expounding
all the relevant theoretical features.
\medskip

Finally, we should add an important remark: since, due to the
universal nature of gravitation, SR should be carefully viewed as an
approximation of GR rather than an \emph{autonomous} theory, great
part of the unending ontological debate about the issues of time,
becoming, endurantism, perdurantism etc.. at the special
relativistic level, should be reconsidered having in view the
results we are going to discuss in the present paper.

\bigskip

In conclusion, here we are only interested in ascertaining whether
and to what extent the notion of instantaneous 3-space is physically
consistent and univocally definable. With this in view, while
sketching the problem of distant simultaneity in the Newtonian and
special-relativistic cases for the sake of argument, we will focus
on the new and unexpected result concerning the natural,
\emph{dynamically ruled}, emergence of a notion of instantaneous
3-space in certain classes of models of GR.

\vfill\eject

\section{Newton's Absolute Distant-Simultaneity.}

The absolute space of Newton is, by definition, an instantaneous
3-space at every value of absolute time. Leaving aside
foundational problems of Newton and Galilei viewpoints, let us
summarize the essential elements of the {\it } physicist's
viewpoint about non-relativistic mechanics.

The arena of Newton physics is Galilei space-time, in which both
time and space have an absolute \footnote{In various senses, the
most important of which is the statement that time and space are
entities independent of the dynamics.} status (its mathematical
structure is the direct product $R^3 \times R$ and can be visualized
as a foliation with base manifold the time axis and with Euclidean
3-spaces as fibers\footnote{In physics it is always assumed that
space, time and space-time can be idealized as suitable mathematical
manifolds possibly with additional structure.}). As a consequence,
we have the absolute notions of simultaneity, instantaneous
Euclidean 3-space and Euclidean spatial distance.

Space is a {\it container} of material bodies, i.e.  objects endowed
only with a  (inertial) mass. The position of an object, unlike
Newton tenets, is a relative frame-dependent notion. Furthermore,
since there is an \emph{absolute temporal distance} between events,
while Newton's instantaneous 3-space is a \emph{metric} space,
Galilei's space-time has a \emph{degenerate metric} structure.

\medskip

{\it Newton's first law}, i.e.  {\it Galilei law of inertia},
states that free objects move on straight lines, eliminating any
intrinsic relevance of  velocity.

{\it Newton's second law}, $\vec F = m\, \vec a$, identifies
acceleration as the basic absolute quantity in the description of
motion, where the force is intended to be measured statically.
\medskip

{\it Galilei relativity principle} selects the inertial frames
centered on inertial observers as the preferred ones due to the
form-invariance of the second law under the Galilei
transformations connecting inertial frames.

\medskip

Gravity is described by an instantaneous action-at-a-distance
interaction enjoying the special property of the equality of
inertial and gravitational masses ({\it Galilei equivalence
principle}).\bigskip

In this absolutist point of view, the absolute existence of
3-space allows to develop a well posed kinematics of isolated
point-like N-body systems, which can be extended to rigid bodies
(and then extended also to deformable ones, see for instance
molecular physics). Given the  positions ${\vec x}_i(t)$,
$i-1,..,N$, of the bodies of mass $m_i$ in a given inertial frame,
we can uniquely define their {\it center of mass} $\vec x(t) =
\sum_i\, m_i\, {\vec x}_i(t) / \sum_i\, m_i$, which describes a
{\it decoupled pseudo-particle in inertial motion}, ${{d^2\ {\vec
x}(t)}\over {dt^2}} = 0$. All the dynamics is shifted to N-1
relative variables ${\vec r}_a(t)$, $a=1,..,N-1$. At the
Hamiltonian level, where the 3-velocities ${\dot {\vec x}}_i(t) =
{{d {\vec x}_i(t)}\over {dt}}$ are replaced by the momenta ${\vec
p}_i(t) = m_i\, {\dot {\vec x}}_i(t)$, the separation of the
center-of-mass conjugate variables ($\vec x(t)$, $\vec p =
\sum_i\, {\vec p}_i(t) = const.$) from any set of conjugate
relative variables ${\vec r}_a(t)$, ${\vec \pi}_a(t)$, is realized
by a canonical transformation, which is a {\it point}
transformation  in the coordinates and the momenta separately.

\bigskip

The main property of the non-relativistic notion of center of mass
is that it can be determined locally in 3-space in the region
occupied by the particles and does not depend on the complementary
region of 3-space.  Naively, one could say that if we eliminate the
decoupled inertial pseudo-particle describing the center of mass, we
shift to a {\it relational} description based on a set of {\it
relative variables} ${\vec r}_a(t)$, ${\vec \pi}_a(t)$. However, as
shown in Ref.\cite{1}, this is true only if the total barycentric
angular momentum of the non-relativistic universe is {\it zero}
\footnote{This is connected with the fact that, while the 3-momentum
Noether constants of motion satisfy an Abelian algebra at the
Hamiltonian level, the angular momentum Noether constants of motion
satisfy a non-Abelian one (only two of them, ${\vec J}^2$ and $J_3$,
have vanishing Poisson brackets). As shown in Ref.\cite{2} this
explains why we can decouple the center of mass globally. On the
contrary, there is no unique global way to separate 3 rotational
degrees of freedom (see molecular dynamics \cite{3}): this gives
rise to \emph{dynamical body frames} for deformable bodies (think of
the diver and the falling cat). As well known and as shown in
Ref.\cite{4}, in special relativity there is no notion of center of
mass enjoying all the properties of the non-relativistic ones. See
Refs.\cite{4,5} for the types of non-relativistic relative variables
admitting a relativistic extension.}. If the total angular momentum
is different from zero, Newton relative motion in absolute 3-space
satisfies equations of motion which are different from equations of
motion having a purely relational structure \cite{1}. This shows why
the interpretations of the Newton rotating bucket are so different
in the absolute and the relational descriptions.

\medskip

Let us remark that, since inertial observers are idealizations, all
realistic observers are (linearly and/or rotationally) accelerated,
so that their spatial trajectories can be taken as the time axis of
global (rigid or non-rigid) {\it non-inertial} frames. It turns out
that in Newton's theory this leads only to the appearance of
inertial  forces proportional to the inertial mass of the
accelerated body, rightly called fictitious (or apparent). Let us
stress, on the other hand, that all realistic observers do
experience such forces and thereby have the problem of disentangling
the real dynamical forces from the apparent ones.

\vfill\eject

\section{Special Relativity: Conventional Distant Simultaneity}

The arena of special relativity is Minkowski space-time, in which
only the (3+1)-dimensional space-time is an absolute notion. Its
Lorentz signature and its absolute chrono-geometrical structure
allow to distinguish time-like, null and space-like intervals.
However, given the world-line of a time-like observer, in each
point the observer can only identify the conformal structure of
the incoming or outgoing rays of light (the past and future fixed
light-cone in that point): for the observer there is no intrinsic
notion of simultaneous events, of instantaneous 3-space (to be
used as a Cauchy surface for Maxwell equations), of spatial
distance, or 1-way velocity of light. The definition of
instantaneous 3-space is then completely ruled by the conventions
about distant simultaneity.
\bigskip

As well-known, the starting point is constituted by the following
basic postulates:

\noindent Two {\it light postulates} - The 2-way (or round-trip)
velocity of light (only one clock is involved in its definition)
is A) isotropic and B) constant ($= c$).

\noindent The {\it relativity principle} (replacing the Galilei
one) - It selects the relativistic inertial frames, centered on
inertial time-like observers, and the Cartesian 4-coordinates
$x^{\mu}$, in which the line element is $ds^2 = \eta_{\mu\nu}\,
dx^{\mu}\, dx^{\nu}$, $\eta_{\mu\nu} = \sgn\, (+---)$, $\sgn = \pm
1$ (according to particle physics or general relativity
conventions, respectively).
\medskip

The {\it law of inertia} states now that a test body moves along a
flat time-like geodesics (a null one for a ray of light). The
Poincar\'e group defines the transformations among the inertial
frames. The preferred inertial frames are also selected by {\it
Einstein's convention} for the synchronization of the clock of the
inertial observer with any (in general accelerated) distant clock
\footnote{The inertial observer ($\gamma$) sends rays of light to
another time-like observer $\gamma_1$, who reflects them back
towards $\gamma$. Given the emission ($\tau_i$) and adsorption
($\tau_f$) times on $\gamma$, the point $P$ of reflection on
$\gamma_1$ is assumed to be simultaneous with the point $Q$ on
$\gamma$ where $\tau_Q = \tau_i + {1\over 2}\, (\tau_f - \tau_i) =
{1\over 2}\, (\tau_i + \tau_f) {\buildrel {def}\over =}\, \tau_P$.
With this so-called {\it Einstein's ${1\over 2}$ convention for the
synchronization of distant clocks}, the {\it instantaneous 3-space}
is the space-like hyper-plane $x^o = const.$ {\it orthogonal to
$\gamma$}, the point $Q$ is the midpoint between the emission and
adsorption points and, since $\tau_P - \tau_i = \tau_f - \tau_P$,
the {\it one-way} velocity of light between $\gamma$ and every
$\gamma_1$ is isotropic and equal to the round-trip velocity of
light $c$. Note that in modern metrology clock synchronization is
always performed by means of light signals and the velocity of light
is assumed as a standard replacing the traditional unit of length.},
according to which the inertial instantaneous 3-spaces are the
Euclidean space-like hyper-planes $x^o = c\, t = const.$ Only with
this convention the 1-way velocity of light between the inertial
observer and any accelerated one coincides with the 2-way velocity
$c$.\medskip

The spatial distance between two simultaneous events in an inertial
frame is the Euclidean distance along the connecting flat
3-geodesics.

\bigskip

Since inertial frames are still an idealization, we must consider
the non-inertial ones centered on accelerated observers. As shown in
Refs.\cite{6} (see also Refs.\cite{7,8}) the standard 1+3 approach,
trying to build non-inertial coordinates starting from the
world-line of the accelerated observers, meets coordinate
singularities preventing their global definition \footnote{Fermi
coordinates, defined on hyper-planes orthogonal to the observer's
4-velocity become singular where the hyper-planes intersect, i.e. at
distances from the world-line of the order of the so-called linear
and rotational {\it acceleration radii} (${\cal L} = {{c^2}\over a}$
for an observer with translational acceleration $a$; ${\cal L} =
{c\over {\Omega}}$ for an observer rotating with frequency $\Omega$)
\cite{6,9} (see also Ref.\cite{10}). For rotating coordinates
(rotating disk with the associated Sagnac effect) there is a
coordinate singularity (the component $g_{oo}$ of the associated
4-metric vanishes) at a distance from the rotation axis, where the
tangential velocity becomes equal to $c$ (the so-called horizon
problem) \cite{6}.}. Let us remember that the theory of measurements
in non-inertial frames is based on the {\it locality principle}
\cite{9}: standard clocks and rods do not feel acceleration and at
each instant the detectors of the instantaneously comoving inertial
observer give the correct data. Again this procedure fails in
presence of electro-magnetic fields when their wavelength is of the
order of the acceleration radii \cite{9} (the observer is not static
enough during 5-10 cycles of such waves, so that their frequency
cannot be measured).\bigskip

The only known method to overcome these difficulties is to shift to
the 3+1 point of view, in which, given the world-line of the
observer, one adds as an {\it independent structure} a 3+1 splitting
of Minkowski space-time, which is nothing else than a clock
synchronization convention. This allows to define a global
non-inertial frame centered on the observer. This splitting foliates
Minkowski space-time with space-like hyper-surfaces $\Sigma_{\tau}$,
which are the instantaneous (Riemannian) 3-spaces \footnote{Let us
stress that each instantaneous 3-space is a possible Cauchy surface
for Maxwell equations. Namely the added structure allows to have a
well-posed initial value problem for these equations and to apply to
them the theorem on the existence and uniqueness of the solutions of
partial differential equations. The price to guarantee {\it
predictability} is the necessity of giving Cauchy data on a
non-compact space-like 3-surface inside Minkowski space-time. This
is the unavoidable element of {\it non-factuality} which the 1+3
point of view would like to avoid.} associated to the given
convention for clock synchronization (in general different from
Einstein's). The leaves $\Sigma_{\tau}$ of the foliation are labeled
by any scalar monotonically increasing function of the proper time
of the observer. The intersection point of the observer world-line
with each $\Sigma_{\tau}$ is chosen as the origin of scalar
curvilinear 3-coordinates. Such observer-dependent 4-coordinates
$\sigma^A = (\tau ; \sigma^r)$ are called {\it radar 4-coordinates}.
Now the 1-way velocity of light becomes, in general, both
anisotropic and point-dependent, while the (Riemannian) spatial
distance between two simultaneous points on $\Sigma_{\tau}$ is
defined along the 3-geodesic joining them.\medskip

If $x^{\mu} = z^{\mu}(\tau ,\vec \sigma )$ describes the embedding
of the associated simultaneity 3-surfaces $\Sigma_{\tau}$ into
Minkowski space-time, so that the metric induced by the coordinate
transformation $x^{\mu} \mapsto \sigma^A = (\tau ,\vec \sigma )$
is $g_{AB}(\tau ,\vec \sigma ) = {{\partial z^{\mu}(\sigma )}\over
{\partial \sigma^A}} \, \eta_{\mu\nu}\, {{\partial z^{\nu}(\sigma
 )}\over {\partial \sigma^B}}$ \footnote{The 4-vectors $z^{\mu}_r(\tau ,\vec \sigma )
= {{\partial z^{\mu}(\tau ,\vec \sigma )}\over {\partial
\sigma^r}}$ are tangent to $\Sigma_{\tau}$. If $l^{\mu}(\tau ,\vec
\sigma )$ is the unit normal to $\Sigma_{\tau}$ (proportional to
$\epsilon^{\mu}{}_{\alpha\beta\gamma}\, [z^{\alpha}_1\,
z^{\beta}_2\, z^{\gamma}_3](\tau ,\vec \sigma )$), we have
$z^{\mu}_{\tau}(\tau ,\vec \sigma ) = {{\partial z^{\mu}(\tau
,\vec \sigma )}\over {\partial \tau}} = N(\tau ,\vec \sigma )\,
l^{\mu}(\tau ,\vec \sigma ) + N^r(\tau ,\vec \sigma )\,
z^{\mu}_r(\tau ,\vec \sigma )$, where $N(\tau ,\vec \sigma )$ and
$N^r(\tau ,\vec \sigma )$ are the lapse and shift functions,
respectively.}, the basic restrictions on the 3+1 splitting
(leading to a nice foliation with space-like leaves) are the
M$\o$ller conditions \cite{6}

\bea
 && \sgn\, g_{\tau\tau}(\sigma ) > 0,\nonumber \\
 &&{}\nonumber \\
 && \sgn\, g_{rr}(\sigma ) < 0,\qquad \begin{array}{|ll|} g_{rr}(\sigma )
 & g_{rs}(\sigma ) \\ g_{sr}(\sigma ) & g_{ss}(\sigma ) \end{array}\, > 0, \qquad
 \sgn\, det\, [g_{rs}(\sigma )]\, < 0,\nonumber \\
 &&{}\nonumber \\
 &&\Rightarrow det\, [g_{AB}(\sigma  )]\, < 0.
 \label{b1}
 \eea

\medskip

Furthermore, in order to avoid possible asymptotic degeneracies of
the foliation, we must make the additional requirement that the
simultaneity 3-surfaces $\Sigma_{\tau}$ must tend to space-like
hyper-planes at spatial infinity: $z^{\mu}(\tau ,\vec \sigma )
{\rightarrow}_{|\vec \sigma | \rightarrow \infty}\,\, x^{\mu}_s(\tau
) + \epsilon^{\mu}_r\, \sigma^r$ and $g_{AB}(\tau ,\vec \sigma )
{\rightarrow}_{|\vec \sigma | \rightarrow \infty}\,\, \eta_{AB}$,
with the $\epsilon^{\mu}_r$'s being 3 unit space-like 4-vectors
tangent to the asymptotic hyper-plane, whose unit normal is
$\epsilon^{\mu}_{\tau}$ [the $\epsilon^{\mu}_A$ form an asymptotic
cotetrad, $\epsilon^{\mu}_A\, \eta^{AB}\, \epsilon^{\nu}_B =
\eta^{\mu\nu}$].

\medskip

As shown in Refs.\cite{6,7}, Eqs.(\ref{b1}) forbid rigid
rotations: {\it only differential rotations} are allowed
(consistently with the modern description of rotating stars in
astrophysics) and the simplest example is given by those 3+1
splittings whose simultaneity 3-surfaces are hyper-planes with
rotating 3-coordinates described by the embeddings ($\sigma =
|\vec \sigma |$)

\bea
 &&z^{\mu}(\tau ,\vec \sigma ) = x^{\mu}(\tau ) + \epsilon^{\mu}_r\,
R^r{}_s(\tau , \sigma )\, \sigma^s,\nonumber \\
 &&{}\nonumber \\
 &&R^r{}_s(\tau ,\sigma ) {\rightarrow}_{\sigma \rightarrow
 \infty} \delta^r_s,\qquad \partial_A\, R^r{}_s(\tau
 ,\sigma )\, {\rightarrow}_{\sigma \rightarrow
 \infty}\, 0,\nonumber \\
 &&{}\nonumber \\
 &&R(\tau ,\sigma ) = \tilde R(\beta_a(\tau ,\sigma )),\qquad
\beta_a(\tau ,\sigma ) = F(\sigma )\, {\tilde \beta}_a(\tau ),
\quad a=1,2,3,\nonumber \\
 &&{}\nonumber \\
 &&{{d F(\sigma )}\over {d \sigma}} \not= 0,\qquad 0 < F(\sigma )
< {1\over {A\, \sigma}}.
 \label{b2}
 \eea

Each $F(\sigma )$ satisfying the restrictions of the last line,
coming from Eqs.(\ref{b1}), gives rise to a {\it global
differentially rotating non-inertial frame}.
\bigskip

Since physical results in special relativity must not depend on
the clock synchronization convention, a description including both
standard inertial frames and admissible non-inertial ones is
needed. This led to the discovery of {\it parametrized Minkowski
theories}.\medskip

As shown in Refs.\cite{11} (see also Refs.\cite{6,7,8}), given the
Lagrangian of every isolated system, one makes the coupling to an
external gravitational field and then replaces the external metric
with the $g_{AB}(\tau ,\vec \sigma )$ associated to a
M$\o$ller-admissible 3+1 splitting. The resulting action principle
$S = \int d\tau d^3\sigma\, {\cal L}(matter , g_{AB}(\tau ,\vec
\sigma ))$ depends upon the system and the embedding $z^{\mu}(\tau
,\vec \sigma )$ and is invariant under frame-preserving
diffeomorphisms: $\tau \mapsto \tau^{'}(\tau ,\vec \sigma  )$,
$\sigma^r \mapsto \sigma^{{'}\, r}(\vec \sigma )$. This {\it
special-relativistic general covariance} implies the vanishing of
the canonical Hamiltonian and the following 4 first class
constraints

\bea
 {\cal H}_{\mu}(\tau ,\vec \sigma  ) &=& \rho_{\mu}(\tau ,\vec
\sigma ) - \sgn\, l_{\mu}(\tau ,\vec \sigma )\, {\cal M}(\tau
,\vec \sigma ) - \sgn\, z_{r \mu}(\tau ,\vec \sigma )\,
h^{rs}(\tau ,\vec \sigma )\, {\cal M}_s(\tau ,\vec \sigma )
\approx 0,\nonumber \\
&&\{ {\cal H}_{\mu}(\tau ,\vec \sigma ), {\cal H}_{\nu}(\tau
,{\vec \sigma}^{'}) \} = 0,
 \label{b3}
\eea

\noindent where $\rho_{\mu}(\tau ,\vec \sigma )$ is the  momentum
conjugate to $z^{\mu}(\tau ,\vec \sigma )$ and $[\sum_u\, h^{ru}\,
g_{us}](\tau ,\vec \sigma ) = \delta^r_s$. ${\cal M}(\tau ,\vec
\sigma ) = T_{\tau\tau}(\tau ,\vec \sigma )$ and ${\cal M}_r(\tau
,\vec \sigma ) = T_{\tau r}(\tau ,\vec \sigma )$ are the energy-
and momentum- densities of the isolated system in
$\Sigma_{\tau}$-adapted coordinates [for N free particles we have
${\cal M}(\tau ,\vec \sigma ) = \sum_{i=1}^N\, \delta^3(\vec
\sigma - {\vec \eta}_i(\tau ))\, \sqrt{m^2_i + h^{rs}(\tau ,\vec
\sigma )\, \kappa_{ir}(\tau )\, \kappa_{is}(\tau )}$, ${\cal
M}_r(\tau ,\vec \sigma ) = \sum_{i=1}^N\, \delta^3(\vec \sigma -
{\vec \eta}_i(\tau ))\, \kappa_{ir}(\tau )$].

Since the matter variables have  only $\Sigma_{\tau}$-adapted
Lorentz-scalar indices, the 10 constant of the motion corresponding
to the generators of the external Poincar\'e algebra are

\bea
 P^{\mu} &=& \int d^3\sigma\, \rho^{\mu}(\tau ,\vec \sigma
),\nonumber \\
 J^{\mu\nu} &=& \int d^3\sigma\, [z^{\mu}\, \rho^{\nu} -
z^{\nu}\, \rho^{\mu}](\tau ,\vec \sigma  ).
 \label{b4}
\eea

\medskip

The Hamiltonian gauge transformations generated by constraints
(\ref{b3}) change the form and the coordinatization of the
simultaneity 3-surfaces $\Sigma_{\tau}$: as a consequence, the
embeddings $z^{\mu}(\tau ,\vec \sigma )$ are {\it gauge variables},
so that in this framework the choice of the non-inertial frame and
in particular of the convention for the synchronization of distant
clocks \cite{6,7} is a {\it gauge choice}. All the inertial and
non-inertial frames compatible with the M$\o$ller conditions
(\ref{b1}) are {\it gauge equivalent} for the description of the
dynamics of isolated systems.
\bigskip

A subclass of the embeddings $z^{\mu}(\tau ,\vec \sigma )$, in
which the simultaneity leaves $\Sigma_{\tau}$ are equally spaced
hyper-planes, describes the standard inertial frames if an
inertial observer is chosen as origin of the 3-coordinates. Let us
stress that every isolated system intrinsically identifies a
special inertial frame, i.e. the rest frame. The use of radar
coordinates in the rest frame leads to parametrize the dynamics
according to the {\it Wigner-covariant rest-frame instant form of
dynamics} developed in Refs.\cite{11}. This instant form is a
special case of {\it parametrized Minkowski theories} \cite{11}
\cite{7} \footnote{This approach was developed to give a
formulation of the N-body problem on arbitrary simultaneity
3-surfaces. The change of clock synchronization convention may be
formulated as a {\it gauge transformation} not altering the
physics and there is no problem in introducing the
electro-magnetic field when the particles are charged (see
Refs.\cite{11,12,13}). The rest-frame instant form corresponds to
the gauge choice of the 3+1 splitting whose simultaneity
3-surfaces are the intrinsic rest frame of the given configuration
of the isolated system. See Ref.\cite{14} for the Hamiltonian
treatment of the relativistic center-of-mass problem and for the
issue of reconstructing orbits in the 2-body case.}, in which the
leaves of the 3+1 splitting of Minkowski space-time are inertial
hyper-planes (simultaneity 3-surfaces called {\it Wigner
hyper-planes}) orthogonal to the conserved 4-momentum $P^{\mu}$ of
the isolated system.
\bigskip

The inertial rest-frame instant form is associated with the special
gauge $z^{\mu}(\tau ,\vec \sigma ) = x^{\mu}_s(\tau ) +
\epsilon^{\mu}_r(u(P))\, \sigma^r$, $x^{\mu}_s(\tau ) =
Y^{\mu}_s(\tau ) = u^{\mu}(P)\, \tau$, selecting the inertial rest
frame of the isolated system centered on the Fokker-Pryce 4-center
of inertia and having as instantaneous 3-spaces the Wigner
hyper-planes.

\bigskip

Another particularly interesting family of 3+1 splittings of
Minkowski space-time is defined by the embeddings

\bea
 z^{\mu}(\tau ,\vec \sigma ) &=& Y^{\mu}_s(\tau ) + F^{\mu}(\tau
,\vec \sigma ) = u^{\mu}(P)\, \tau + F^{\mu}(\tau ,\vec \sigma
),\qquad F^{\mu}(\tau , \vec 0) = 0,\nonumber \\
 &&{\rightarrow}_{\sigma \rightarrow \infty}\, u^{\mu}(P)\, \tau +
\epsilon^{\mu}_r(u(P))\, \sigma^r,
 \label{b5}
\eea

\noindent with $F^{\mu}(\tau ,\vec \sigma )$ satisfying
Eqs.(\ref{b1}).
\medskip

In this family the simultaneity 3-surfaces $\Sigma_{\tau}$ tend to
Wigner hyper-planes at spatial infinity, where they are orthogonal
to the conserved 4-momentum of the isolated system. Consequently,
there are asymptotic inertial observers with world-lines parallel to
that of the Fokker-Pryce 4-center of inertia, namely there are the
{\it rest-frame conditions} $p_r = \epsilon^{\mu}_r(u(P))\, P_{\mu}
= 0$, so that the embeddings (\ref{b5}) define {\it global
M$\o$ller-admissible non-inertial rest frames} \footnote{The only
ones existing in tetrad gravity, due to the equivalence principle,
in globally hyperbolic asymptotically flat space-times without
super-translations as we shall see in the next Section.}.
\medskip

Since we are in non-inertial rest frames, the internal energy- and
boost- densities contain the inertial potentials source of the
relativistic inertial forces (see Ref.\cite{15} for the quantization
in non-inertial frames): more precisely, they are contained in the
spatial components of the metric $g_{rs}(\tau ,\vec \sigma )$
associated to the embeddings (\ref{b5}).
\medskip

\bigskip

In conclusion {\it the only notion of instantaneous 3-space} which
can be introduced in special relativity is always observer $
[X^{\mu}(\tau)$]- and frame [$\Sigma_{\tau}$]-dependent. The
conceptual difficulties connected with the notion of relativistic
center of mass also show that its definition (using the global
Poincar\'e generators of the isolated system) \emph{necessitates a
whole instantaneous 3-space }$\Sigma_{\tau}$. Therefore, even if we
eventually get a decoupled pseudo-particle like in Newton theory
\footnote{However, the canonical transformations decoupling it from
the relative variables are now non-point; only for free particles
they remain point in the momenta, but not in the positions
\cite{14}.}, we lose the possibility of treating disjoint,
non-interacting, sub-systems independently of one another. This is
just due to the necessity of choosing a convention for the
synchronization of distant clocks.
\bigskip

\vfill\eject

\section{General Relativity: Dynamically Determined Distant Simultaneity}

We will show that, contrary to a widespread opinion, general
relativity - at least for a particular class of models
\footnote{Given the enormous variety of solutions of Einstein's
equations, one cannot expect to find general answers to ontological
questions.} - contains in itself the capacity for a dynamical
definition of instantaneous 3-spaces. Every model of GR in the given
class, once completely specified in a precise sense that will be
explained presently, entails that space-time \emph{be } essentially
the unfolding in time of a dynamically variable form of
instantaneous 3-space.

\bigskip

In the years 1913-16 Einstein developed general relativity by
relying on the {\it equivalence principle} (equality of inertial and
gravitational masses of non-spinning test bodies in free fall) and
on the guiding principle of general covariance. Einstein's original
view was that the principle had to express the impossibility of
distinguishing a uniform gravitational field from the effects of a
constant acceleration by means of local experiments in a
sufficiently small region with negligible tidal forces. This led him
to the {\it geometrization} of the gravitational interaction and to
the replacement of Minkowski space-time with a pseudo-Riemannian
4-manifold \emph{$M^4$} with non vanishing curvature Riemann tensor.
\medskip

The \emph{equivalence principle} entails the non existence of global
inertial frames (SR relativity holds only in a small neighbourhood
of a body in free fall).  The principle of {\it general covariance}
(see Ref.\cite{16} for a thorough review), which expresses the
tensorial nature of Einstein's equations, has the following two
consequences:

i) the invariance of the Hilbert action under {\it passive}
diffeomorphisms (the coordinate transformations in \emph{$M^4$}), so
that the second Noether theorem implies the existence of first-class
constraints at the Hamiltonian level;

ii) the mapping of the solutions of Einstein's equations among
themselves under the action of {\it active} diffeomorphisms of
\emph{$M^4$} extended to the tensors over \emph{$M^4$
}(\emph{dynamical symmetries} of Einstein's equations).

\bigskip

The basic field of metric gravity is the 4-metric tensor with
components ${}^4g_{\mu\nu}(x)$ in an arbitrary coordinate system of
\emph{$M^4$.} The peculiarity of gravity is that the 4-metric field,
unlike the fields of electromagnetic, weak and strong interactions
and the matter fields, has a {\it double role}:

i) it is the mediator of the gravitational interaction (in analogy
to all of the other gauge fields);

ii) it determines \emph{dynamically} the chrono-geometric structure
of space-time \emph{$M^4$} through the line element $ds^2 =
{}^4g_{\mu\nu}(x)\, dx^{\mu}\, dx^{\nu}$.

Consequently, the gravitational field {\it teaches relativistic
causality} to all of the other fields: in particular classical rays
of light, photons and gluons, which are the trajectories allowed for
massless particles in each point of \emph{$M^4$.}

\bigskip

Let us make a comment about the two main existing approaches for
quantizing gravity.

\medskip

1) {\it Effective quantum field theory and string theory}. This
approach contains the standard model of elementary particles and its
extensions. However, since the quantization, namely the definition
of the Fock space, requires a background space-time for the
definition of creation and annihilation operators, one must use the
splitting ${}^4g_{\mu\nu} = {}^4\eta^{(B)}_{\mu\nu} +
{}^4h_{\mu\nu}$ and quantize only the perturbation ${}^4h_{\mu\nu}$
of the background 4-metric $\eta^{(B)}_{\mu\nu}$ (usually $B$ is
either Minkowski or DeSitter space-time). In this way property ii)
is lost (one exploits the fixed non-dynamical chrono-geometrical
structure of the background space-time), and gravity is replaced by
a field of spin two over the background (and passive diffeomorphisms
are replaced by a Lie group of gauge transformations acting in an
"inner" space). The only difference between gravitons, photons and
gluons lies thereby in their quantum numbers.

\medskip

2) {\it Loop quantum gravity}. This approach does not introduce a
background space-time but, being inequivalent to a Fock space, has
problems in incorporating particle physics. It exploits a fixed 3+1
splitting of the space-time \emph{$M^4$} and quantizes the
associated instantaneous 3-spaces $\Sigma_{\tau}$ (quantum
geometry). There is no known way, however, to implement consistent
unitary evolution (the problem of the super-Hamiltonian constraint).
Furthermore, since the theory is usually formulated in
spatially-compact space-times without boundary, it admits no
Poincar\'e symmetry group (and therefore no extra-dimensions as in
string theory), and faces a serious problem concerning the
definition of time: the so-called {\it frozen picture} without real
evolution.

For outside points of view on loop quantum gravity and string theory
see Ref.\cite{17,18}, respectively.

\bigskip

Let us remark that all formulations of the theory of elementary
particle and nuclear physics are a chapter of the theory of
representations of the Poincar\'e group in the {\it inertial frames}
of the spatially non-compact Minkowski space-time. As a consequence,
if one looks at general relativity from the point of view of
particle physics, the main problem to get a unified theory is that
of conciliating the Poincar\'e group and the diffeomorphism group.

\bigskip

Let us now consider the ADM formulation of metric gravity \cite{19}
and its extension to tetrad gravity \footnote{This extension is
needed to describe the coupling of gravity to fermions; it is a
theory of time-like observers each one endowed with a tetrad field,
whose time-like axis is the unit 4-velocity of the observer and
whose spatial axes are associated with a choice of three
gyroscopes.} obtained by replacing the ten configurational 4-metric
variables ${}^4g_{\mu\nu}(x)$ with the sixteen cotetrad fields
${}^4E^{(\alpha )}_{\mu}(x)$ by means of the decomposition
${}^4g_{\mu\nu}(x) = {}^4E^{(\alpha )}_{\mu}(x)\, {}^4\eta_{(\alpha
)(\beta )}\, {}^4E^{(\beta )}_{\nu}(x)$ [$(\alpha )$ are flat
indices].\medskip

Then, after having restricted the model to globally-hyperbolic,
topologically trivial, spatially non-compact space-times (admitting
a {\it global notion  of time}), let us introduce a global 3+1
splitting of the space-time $M^4$ and choose the world-line of a
time-like observer. As in special relativity, let us make a
coordinate transformation to observer-dependent {\it radar}
4-coordinates, $x^{\mu} \mapsto \sigma^A = (\tau ,\sigma^r)$,
adapted to the 3+1 splitting and using the observer world-line as
origin of the 3-coordinates. Again, the inverse transformation,
$\sigma^A \mapsto x^{\mu} = z^{\mu}(\tau ,\sigma^r)$, defines the
embedding of the leaves $\Sigma_{\tau}$ into $M^4$. These leaves
$\Sigma_{\tau}$ (assumed to be Riemannian 3-manifolds diffeomorphic
to $R^3$, so that they admit global 3-coordinates $\sigma^r$ and a
unique 3-geodesic joining any pair of points in $\Sigma_{\tau}$) are
both Cauchy surfaces and \emph{simultaneity surfaces }corresponding
to a \emph{convention for clock synchronization}. For the induced
4-metric we get

\begin{eqnarray*}
 {}^4g_{AB}(\sigma ) &=& {{\partial z^{\mu}(\sigma )}\over
{\partial \sigma^A}}\, {}^4g_{\mu\nu}(x)\, {{\partial z^{\nu}(\sigma
)}\over {\partial \sigma^B}} =
 {}^4E^{(\alpha )}_A\, {}^4\eta_{(\alpha )(\beta )}\,
{}^4E^{(\beta )}_B = \nonumber \\
 &=&\epsilon \left( \begin{array}{cc} (N^2- {}^3g_{rs}\, N^r\, N^s) &
- {}^3g_{su}\, N^u\\ - {}^3g_{ru}\, N^u & -{}^3g _{rs} \end{array}
\right)(\sigma ).\nonumber \\
 \end{eqnarray*}

Here ${}^4E^{(\alpha )}_A(\tau ,\sigma^r)$ are adapted cotetrad
fields, $N(\tau ,\sigma^r)$ and $N^r(\tau ,\sigma^r)$ the lapse and
shift functions, and ${}^3g_{rs}(\tau ,\sigma^r)$ the 3-metric on
$\Sigma_{\tau}$ with signature $(+ + +)$. We see that, unlike in
special relativity, in general relativity the quantities $z^{\mu}_A
= \partial z^{\mu}/\partial \sigma^A$  are no more cotetrad fields
on $M^4$. Here they are only transition functions between coordinate
charts, so that the dynamical fields are now the real cotetrad
fields ${}^4E^{(\alpha )}_A(\tau ,\sigma^r)$ and not the embeddings
$z^{\mu}(\tau ,\sigma^r)$.

\bigskip

Let us try to identify a class of space-times and an associated
family of admissible 3+1 splittings suitable to incorporate particle
physics and provide a model for the solar system or our galaxy (and
hopefully even allowing an extension to the cosmological context)
with the following further requirements \cite{20}:
\medskip

1) $M^4$ must be asymptotically flat at spatial infinity and the
4-metric must tend asymptotically to the Minkowski 4-metric there,
in every coordinate system (this implies that the 4-diffeomorphisms
must tend there to the identity at spatial infinity). In such
space-times, therefore, there is an {\it asymptotic background
4-metric} allowing to avoid the decomposition ${}^4g_{\mu\nu} =
{}^4\eta_{\mu\nu} + {}^4h_{\mu\nu}$ in the bulk.

\medskip

2) The boundary conditions on the fields on each leaf
$\Sigma_{\tau}$ of the admissible 3+1 splittings must be such to
reduce the \emph{Spi} group of asymptotic symmetries (see
Ref.\cite{21}) to the ADM Poincar\'e group. This means that there
should not be {\it super-translations} (direction-dependent quasi
Killing vectors, obstruction to the definition of angular momentum
in general relativity), namely that all the fields must tend to
their asymptotic limits in a direction- independent way (see Refs.
\cite{22}). This is possible only if the admissible 3+1 splittings
have all the leaves $\Sigma_{\tau}$ tending to Minkowski space-like
hyper-planes orthogonal to the ADM 4-momentum at spatial infinity
\cite{20}. In turn this implies that every $\Sigma_{\tau}$ is {\it
the rest frame of the instantaneous 3-universe} and that there are
asymptotic inertial observers to be identified with the {\it fixed
stars}\footnote{In a final extension to the cosmological context
they could be identified with the privileged observers at rest with
respect to the background cosmic radiation.}. This requirement
implies that the shift functions vanish at spatial infinity
[$N^r(\tau ,\sigma^r)\, \rightarrow O(1/|\sigma |^\epsilon )$,
$\epsilon > 0$, $\sigma^r = |\sigma |\, {\hat u}^r$], where the
lapse function tends to $1$ [$N(\tau ,\sigma^r)\, \rightarrow\, 1 +
O(1/|\sigma |^\epsilon )$] and the 3-metric tends to the Euclidean
3-metric [${}^3g_{rs}(\tau ,\sigma^u)\, \rightarrow\, \delta_{rs} +
O(1/|\sigma |)$].
\medskip

3) The admissible 3+1 splittings should have the leaves
$\Sigma_{\tau}$ admitting a generalized Fourier transform (namely
they should be Lichnerowicz 3-manifolds \cite{23} with involution).
This would allow the definition of instantaneous Fock spaces in a
future attempt of quantization.

\medskip

4) All the fields on $\Sigma_{\tau}$ should belong to suitable
weighted Sobolev spaces, so that $M^4$ has no Killing vectors and
Yang-Mills fields on $\Sigma_{\tau}$ do not present Gribov
ambiguities (due to the presence of gauge symmetries and gauge
copies) \cite{24}.
\bigskip

In absence of matter the Christodoulou and Klainermann \cite{25}
space-times are good candidates: they are near Minkowski
space-time in a norm sense, avoid singularity theorems by relaxing
the requirement of conformal completability (so that it is
possible to follow solutions of Einstein's equations on long
times) and admit gravitational radiation at null infinity.

\bigskip

Since the simultaneity leaves $\Sigma_{\tau}$ are the rest frame of
the instantaneous 3-universe, at the Hamiltonian level it is
possible to define {\it the rest-frame instant form of metric and
tetrad gravity} \cite{20,26}. If matter is present, the limit of
this description for vanishing Newton constant will be reduced to
the rest-frame instant form description of the same matter in the
framework of parametrized Minkowski theories, while the ADM
Poincar\'e generators will tend to the kinematical Poincar\'e
generators of special relativity. In this way we get a model
admitting {\it a deparametrization of general relativity to special
relativity}. It is not known whether the rest-frame condition can be
relaxed in general relativity without bringing in
super-translations, since the answer to this question is connected
with the non-trivial problem of boosts in general relativity.

\bigskip

Let us now come back to ADM tetrad gravity. The time-like vector
${}^4E^A_{(o)}(\tau ,\sigma^r)$ of the tetrad field
${}^4E^A_{(\alpha )}(\tau ,\sigma^r)$, dual to the cotetrad field
${}^4E^{(\alpha )}_A(\tau ,\sigma^r)$, may be rotated to become the
unit normal to $\Sigma_{\tau}$ in each point by means of a standard
Wigner boost for time-like Poincar\'e orbits depending on three
parameters $\varphi_{(a)}(\tau ,\sigma^r)$, $a = 1,2,3$:
${}^4E^A_{(o)}(\tau ,\sigma^r) = L^A{}_B(\varphi_{(a)}(\tau
,\sigma^r))\, {}^4{\check E}^B_{(o)}(\tau ,\sigma^r)$. This allows
to define the following cotetrads adapted to the 3+1 splitting (the
so-called {\it Schwinger time gauge}) ${}^4{\check E}^{(o)}_A(\tau
,\sigma^r) = \Big(N(\tau ,\sigma^r); 0\Big)$, ${}^4{\check
E}^{(a)}_A(\tau ,\sigma^r) = \Big(N_{(a)}(\tau ,\sigma^r);
{}^3e_{(a)r}(\tau ,\sigma^r)\Big)$, where ${}^3e_{(a)r}(\tau
,\sigma^r)$ are cotriads fields on $\Sigma_{\tau}$ (tending to
$\delta_{(a)r} + O(1/|\sigma |)$ at spatial infinity) and $N_{(a)} =
N^r\, {}^3e_{(a)r}$. As a consequence, the sixteen cotetrad fields
may be replaced by the fields $\varphi_{(a)}(\tau ,\sigma^r)$,
$N(\tau ,\sigma^r)$, $N_{(a)}(\tau ,\sigma^r)$, ${}^3e_{(a)r}(\tau
,\sigma^r)$, whose conjugate canonical momenta will be denoted as
$\pi_N(\tau ,\sigma^r)$, $\pi_{\vec N\, (a)}(\tau ,\sigma^r)$,
$\pi_{\vec \varphi\, (a)}(\tau ,\sigma^r)$, ${}^3\pi^r_{(a)}(\tau
,\sigma^r)$.

\bigskip

The local invariance of the ADM action entails the existence of 14
first-class constraints (10 primary and 4 secondary):

i) $\pi_N(\tau ,\sigma^r) \approx 0$ implying the secondary
super-Hamiltonian constraint ${\cal H}(\tau ,\sigma^r) \approx 0$;

ii) $\pi_{\vec N\, (a)}(\tau ,\sigma^r) \approx 0$ implying the
secondary super-momentum constraints ${\cal H}_{(a)}(\tau ,\sigma^r)
\approx 0$;

iii) $\pi_{\vec \varphi\, (a)}(\tau ,\sigma^r) \approx 0$;

iv) three constraints $M_{(a)}(\tau ,\sigma^r) \approx 0$
generating rotations of the cotriads.\medskip

Consequently, there are 14 gauge variables, which, as shown in
Refs.\cite{27}, describe the {\it generalized inertial effects} in
the non-inertial frame defined by the chosen admissible 3+1
splitting of $M^4$ centered on an arbitrary time-like observer. The
remaining independent "two + two" degrees of freedom are the gauge
invariant DOs of the gravitational field describing {\it generalized
tidal effects} (see Refs.\cite{27}). The same degrees of freedom
emerge in ADM metric gravity, where the configuration variables $N$,
$N^r$, ${}^4g_{rs}$,  with conjugate momenta $\pi_N$, $\pi_{\vec N\,
r}$, ${}^3\Pi^{rs}$, are restricted by 8 first-class constraints
($\pi_N(\tau ,\sigma^r) \approx 0\, \rightarrow {\cal H}(\tau
,\sigma^r) \approx 0$, $\pi_{\vec N\, r}(\tau ,\sigma^r) \approx 0
\, \rightarrow\, {\cal H}^r(\tau ,\sigma^r) \approx 0$).

\bigskip

As already said, the first-class constraints are the generators of
the Hamiltonian gauge transformations, under which the ADM action is
quasi-invariant (second Noether theorem):\medskip

i) The gauge transformations generated by the four primary
constraints $\pi_N(\tau ,\sigma^r) \approx 0$, $\pi_{\vec N\,
(a)}(\tau ,\sigma^r) \approx 0$, modify the lapse and shift
functions, namely how densely the simultaneity surfaces are packed
in $M^4$ and which points have the same 3-coordinates on each
$\Sigma_{\tau}$.

ii) Those generated by the three super-momentum constraints ${\cal
H}_{(a)}(\tau ,\sigma^r) \approx 0$ change the 3-coordinates on
$\Sigma_{\tau}$.

iii) Those generated by the super-Hamiltonian constraint ${\cal
H}(\tau ,\sigma^r) \approx 0$ transform an admissible 3+1 splitting
into another admissible one by realizing a normal deformation of the
simultaneity surfaces $\Sigma_{\tau}$ \cite{34}. As a consequence,
{\it all the conventions about clock synchronization are gauge
equivalent as in special relativity}.

iv) Those generated by $\pi_{\vec \varphi\, (a)}(\tau ,\sigma^r)
\approx 0$, $M_{(a)}(\tau ,\sigma^r) \approx 0$, change the cotetrad
fields with local Lorentz transformations.

\medskip

In the rest-frame instant form of tetrad gravity there are the three
extra first-class constraints $P^r_{ADM} \approx 0$ (vanishing of
the ADM 3-momentum as {\it rest-frame conditions}). They generate
gauge transformations which change the time-like observer whose
world-line is used as origin of the 3-coordinates.

A fundamental technical point, which is of paramount importance for
the physical interpretation, is the possibility of performing a
separation of the gauge variables from the DOs by means of a
Shanmugadhasan canonical transformation \cite{28}.

In Ref.\cite{26} a Shanmugadhasan canonical transformation adapted
to 13 first class constraints (not to the super-Hamiltonian one,
because no one knows how to solve it except in the Post-Newtonian
approximation) has been introduced and exploited to clarify the
interpretation. There are problems, however, when one introduces
matter.
\bigskip

To avoid the above difficulties, a different Shanmugadhasan
canonical transformation, adapted only to 10 constraints but
allowing the addition of any kind of matter to the rest-frame
instant form of tetrad gravity, has been recently found starting
from a new parametrization of the 3-metric \cite{29}.

The basic idea is that the symmetric 3-metric tensor can be
diagonalized with an orthogonal matrix depending on three Euler
angles $\theta^i(\tau ,\vec \sigma )$. The three eigenvalues
$\lambda_i(\tau ,\vec \sigma )$ are then replaced by the conformal
factor $\phi (\tau ,\vec \sigma )$ of the 3-metric and by two
\emph{tidal} variables $R_{\bar a}(\tau ,\vec \sigma )$, $\bar a
=1,2$. The defining equations and the resulting Shanmugadhasan
canonical transformation are:

\bigskip

\begin{eqnarray*}
 {}^3g_{rs} &=& \sum_{uv}\, V_{ru}(\theta^n)\, \lambda_u\, \delta_{uv}\,
 V^T_{vs}(\theta^n) = \nonumber \\
 &=& \sum_a\, \Big(V_{ra}(\theta^n)\, \Lambda^a\Big)\,
 \Big(V_{sa}(\theta^n)\, \Lambda^a\Big) = \sum_a\,
 {}^3{\bar e}_{(a)r}\, {}^3{\bar e}_{(a)s} = \sum_a\, {}^3e_{(a)r}\,
 {}^3e_{(a)s},\nonumber \\
 &&{}\nonumber \\
 \Lambda^a(\tau ,\vec \sigma ) &{\buildrel {def}\over =}&
 \sum_u\, \delta_{au}\, \sqrt{\lambda_u(\tau ,\vec \sigma )}\, =
\phi^2(\tau ,\vec \sigma )\, e^{\sum_{\bar a}\, \gamma_{\bar ab}\,
R_{\bar a}(\tau ,\vec \sigma )}\nonumber \\
  &&\rightarrow_{r
 \rightarrow \infty}\,\, 1 + {M\over {4r}} + {{a^a}\over
 {r^{3/2}}} + O(r^{-3}),
 \end{eqnarray*}

 \begin{eqnarray*}
 {}^3e_{(a)r} &=& R_{(a)(b)}(\alpha_{(c)})\, {}^3{\bar e}_{(b)r},
 \nonumber \\
 &&{}\nonumber \\
 {}^3{\bar e}_{(a)r} &=& \sum_b\, {}^3e_{(b)r}\, R_{(b)(a)}(\alpha_{(e)}) =
 \sum_u\, \sqrt{\lambda_u}\, \delta_{u(a)}\, V^T_{ur}(\theta^n) =
 V_{ra}(\theta^n)\, \Lambda^a,\nonumber \\
 &&{}\nonumber \\
 {}^3{\bar e}^r_{(a)} &=& \sum_u\, {{\delta_{u(a)}}\over {\sqrt{\lambda_u}}}\,
 V_{ru} = {{V_{ra}(\theta^n)}\over {\Lambda^a}},\nonumber \\
 &&{}\nonumber  \\
 &&{}\nonumber \\
 \phi &=& (det\, {}^3g)^{1/12} = ( {}^3e)^{1/6} = {}^3{\bar e}^{1/6} = (\lambda_1\,
\lambda_2\, \lambda_3)^{1/12} = (\Lambda^1\, \Lambda^2\,
\Lambda^3)^{1/6}.\nonumber \\
 &&{}
 \end{eqnarray*}

\begin{eqnarray}
&&\begin{minipage}[t]{3cm}
\begin{tabular}{|l|l|l|l|} \hline
$\varphi_{(a)}$ & $n$ & $n_{(a)}$ & ${}^3e_{(a)r}$ \\ \hline
$\approx 0$ & $\approx 0$ & $  \approx 0 $ & ${}^3{ \pi}^r_{(a)}$
\\ \hline
\end{tabular}
\end{minipage} \hspace{1cm} {\longrightarrow \hspace{.2cm}} \
\begin{minipage}[t]{4 cm}
\begin{tabular}{|ll|ll|l|} \hline
$\varphi_{(a)}$ & $\alpha_{(a)}$ & $n$ & ${\bar n}_{(a)}$ &  ${}^3{\bar e}_{(a)r}$\\
\hline $\approx0$ &
 $\approx 0$ & $\approx 0$ & $\approx 0$
& ${}^3{\tilde {\bar \pi}}^r_{(a)}$ \\ \hline
\end{tabular}
\end{minipage} \nonumber \\
 &&{}\nonumber \\
 &&\hspace{1cm} {\longrightarrow \hspace{.2cm}} \
\begin{minipage}[t]{4 cm}
\begin{tabular}{|ll|ll|ll|} \hline
$\varphi_{(a)}$ & $\alpha_{(a)}$ & $n$ & ${\bar n}_{(a)}$ &
$\theta^r$ & $\Lambda^r$\\ \hline $\approx0$ &
 $\approx 0$ & $\approx 0$ & $\approx 0$
& $\pi^{(\theta )}_r$ & $P_{ r}$ \\ \hline
\end{tabular}
\end{minipage},\nonumber \\
 &&{}\nonumber \\
 &&{}\nonumber \\
  &&\begin{minipage}[t]{3cm}
\begin{tabular}{|l|} \hline
 $\Lambda^r$ \\ \hline $P_r$
\\ \hline
\end{tabular}
\end{minipage} \hspace{1cm} {\longrightarrow \hspace{.2cm}} \
\begin{minipage}[t]{4 cm}
\begin{tabular}{|l|l|} \hline
 $ \phi$ &  $R_{\bar a}$\\
\hline $\pi_{\phi}$ & $\Pi_{\bar a}$ \\
\hline
\end{tabular}
\end{minipage},\nonumber \\
 &&{}\nonumber \\
 &&{}\nonumber\\
  \pi_{\phi} &=& - \sgn\, {{c^3}\over {2\pi\, G}}\,
 ({}^3\bar e)^{5/6}\,\,\, {}^3K = 2\,
 {{\sum_b\, \Lambda^b\, P_b}\over {(\Lambda^1\, \Lambda^2\,
 \Lambda^3)^{1/6}}},
 \end{eqnarray}

 \noindent where ${\bar n}_{(a)} = \sum_b\, n_{(b)}\,
 R_{(b)(a)}(\alpha_{(e)})$ are the shift functions at
 $\alpha_{(a)}(\tau ,\vec \sigma ) = 0$, $\alpha_{(a)}(\tau ,\sigma^r)$
 are three Euler angles and $\theta^r(\tau ,\sigma^r)$ are three angles giving a
coordinatization of the action of 3-diffeomorphisms on the cotriads
${}^3e_{(a)r}(\tau ,\sigma^r)$. The configuration variable $\phi
(\tau ,\sigma^r) = \Big(det\, {}^3g(\tau ,\sigma^r)\Big)^{1/12}$ is
the conformal factor of the 3-metric: it can be shown that it is the
unknown quantity in the super-Hamiltonian constraint (also named the
Lichnerowicz equation). The gauge variables are $n$, ${\bar
n}_{(a)}$, $\varphi_{(a)}$, $\alpha_{(a)}$, $\theta^r$ and
$\pi_{\phi}$, while $R_{\bar a}$, $\Pi_{\bar a}$, $\bar a = 1,2$,
are the DOs of the gravitational field (in general they are not
tensorial quantities).

\bigskip

This canonical transformation is the first explicit construction of
a York map \cite{30}, in which the momentum conjugate to the
conformal factor (\emph{the gauge variable controlling the
convention for clock synchronization}) is proportional to the trace
${}^3K(\tau ,\vec \sigma )$ of the extrinsic curvature of the
simultaneity surfaces $\Sigma_{\tau}$. Both the \emph{tidal} and the
\emph{gauge} variables can be expressed in terms of the original
variables. Moreover, in a family of completely fixed gauges
differing with respect to the convention of clock synchronization,
the deterministic Hamilton equations for the \emph{tidal} variables
and for \emph{matter} variables contain \emph{relativistic inertial
forces} determined by ${}^3K(\tau ,\vec \sigma )$, which change from
attractive to repulsive where the trace changes sign. These inertial
forces do not have a non-relativistic counterpart (the Newton
3-space is absolute) and could perhaps support the proposal of Ref.
\cite{31} \footnote{The model proposed in Ref.\cite{31} is too
naive, as shown by the criticism in Refs.\cite{32}.} according to
which {\it dark matter} could be explained as an {\it inertial
effect}. While in the MOND model \cite{33} there is an arbitrary
function on the acceleration side of Newton equations in the
absolute Euclidean 3-space, here we have the arbitrary gauge
function ${}^3K(\tau ,\vec \sigma )$ on the force side of Hamilton
equations.

\bigskip

Finally let us see which Dirac Hamiltonian $H_D$ generates the
$\tau$-evolution in ADM canonical gravity. In {\it spatially compact
space-times without boundary} $H_D$ is a linear combination of the
primary constraints plus the secondary super-Hamiltonian and
super-momentum constraints multiplied by the lapse and shift
functions, respectively (a consequence of the Legendre transform).
Consequently, $H_D \approx 0$ and, in the reduced phase space, we
get a vanishing Hamiltonian. This implies the so-called {\it frozen
picture} and the problem of how to reintroduce a temporal evolution
\footnote{See Refs.\cite{35} for the problem of time in general
relativity.}. Usually one considers the normal (time-like)
deformation of $\Sigma_{\tau}$ induced by the super-Hamiltonian
constraint as evolution in a local time variable to be identified
(the "multi-fingered" time point of view with a local, either
extrinsic or intrinsic, time): this is the so-called {\it
Wheeler-DeWitt interpretation} \footnote{Kuchar \cite{36} says that
the super-Hamiltonian constraint must not be interpreted as a
generator of gauge transformations, but as an effective
Hamiltonian.}.

\medskip

On the contrary, {\it in spatially non-compact space-times} the
definition of functional derivatives and the existence of a
well-posed Hamiltonian action principle (with the possibility of a
good control of the surface terms coming from integration by
parts) require the addition of the {\it DeWitt surface
term}\cite{37} (living on the surface at spatial infinity) to the
Hamiltonian. It can be shown \cite{20} that in the rest-frame
instant form this term, together with a surface term coming from
the Legendre transformation of the ADM action, leads to the Dirac
Hamiltonian

\begin{equation}
 H_D = {\check E}_{ADM} + (constraints) =
 E_{ADM} + (constraints) \approx E_{ADM}.
\end{equation}

\noindent Here ${\check E}_{ADM}$ is the {\it strong ADM energy}, a
surface term analogous to the one defining the electric charge as
the flux of the electric field through the surface at spatial
infinity in electromagnetism. Since we have ${\check E}_{ADM} =
E_{ADM} + (constraints)$, we see that the non-vanishing part of the
Dirac Hamiltonian is the {\it weak ADM energy} $E_{ADM} = \int
d^3\sigma\, {\cal E}_{ADM}(\tau ,\sigma^r)$, namely the integral
over $\Sigma_{\tau}$ of the ADM energy density (in electromagnetism
this corresponds to the definition of the electric charge as the
volume integral of matter charge density). Therefore there is no
frozen picture but a consistent $\tau$-evolution instead.
 \medskip

Note that the ADM energy density ${\cal E}_{ADM}(\tau ,\sigma^r)$ is
a {\it coordinate-dependent quantity}, because it depends on the
gauge variables (namely upon the relativistic inertial effects
present in the non-inertial frame): this is nothing else than the
old {\it problem of energy} in general relativity. Let us remark
that in most coordinate systems ${\cal E}_{ADM}(\tau ,\sigma^r)$
does not agree with the pseudo-energy density defined in terms of
the Landau-Lifschitz pseudo-tensor.

 \bigskip

In order to get a deterministic evolution for the DOs \footnote{See
Refs.\cite{38} for the modern formulation of the Cauchy problem for
Einstein equations, which mimics the steps of the Hamiltonian
formalism.} we must fix the gauge completely, that is we must add 14
gauge-fixing constraints satisfying an orbit condition and to pass
to Dirac brackets. As already said, the correct way to do so is the
following one:

i) Add a gauge-fixing constraint to the secondary super-Hamiltonian
constraint \footnote{The special choice $\pi_{\phi}(\tau ,\sigma^r)
\approx 0$ implies that the DOs $R_{\bar a}$, $\Pi_{\bar a}$, remain
canonical even if we do not know how to solve this constraint.}.
This gauge-fixing fixes the form of $\Sigma_{\tau}$, i.e. the
convention for the synchronization of clocks. The $\tau$-constancy
of this gauge-fixing constraint generates a gauge-fixing constraint
to the primary constraint $\pi_N(\tau ,\sigma^r) \approx 0$ for the
determination of the lapse function.

ii) Add three gauge-fixings to the secondary super-momentum
constraints ${\cal H}_{(a)}(\tau ,\sigma^r) \approx 0$. This fixes
the 3-coordinates on each $\Sigma_{\tau}$. The $\tau$-constancy of
these gauge fixings generates the three gauge fixings to the primary
constraints $\pi_{\vec N\, (a)}(\tau ,\sigma^r) \approx 0$ and leads
to the determination of the shift functions (i.e. of the appearances
of \emph{gravito-magnetism}).

iii) Add six gauge-fixing constraints to the primary constraints
$\pi_{\vec \varphi\, (a)}(\tau ,\sigma^r) \approx 0$, $M_{(a)}(\tau
,\sigma^r) \approx 0$. This is a fixation of the cotetrad field
which includes a convention on the choice and the transport of the
three gyroscopes of every time-like observer of the two congruences
associated with the chosen 3+1 splitting of $M^4$ (see
Ref.\cite{8,9}.

iv) In the rest-frame instant form we must also add three gauge
fixings to the rest-frame conditions $P^r_{ADM} \approx 0$. The
natural ones are obtained with the requirement that the three ADM
boosts vanish. In this way we select a special time-like observer
as origin of the 3-coordinates (like the Fokker-Pryce center of
inertia in special relativity \cite{5,14}).

\bigskip

In this way all the gauge variables are fixed to be either numerical
functions or well determined functions of the DOs. This complete
gauge fixing is physically equivalent to a definition of the {\it
global non-inertial frame centered on a time-like observer, carrying
its pattern of inertial forces} we have called NIF (see
Ref.\cite{27}). Note that in a NIF, the ADM energy density ${\cal
E}_{ADM}(\tau ,\sigma^r)$ becomes {\it a well defined function of
the DOs only} and the Hamilton equations for them with $E_{ADM}$ as
Hamiltonian are a hyperbolic system of partial differential
equations for their determination. For each choice of Cauchy data
for the DOs on a $\Sigma_{\tau}$, we obtain a solution of Einstein's
equations (an Einstein universe) in the radar 4-coordinate system
associated with the chosen 3+1 splitting of $M^4$.

Actually, the Cauchy data are the 3-geometry and matter variables on
the Cauchy 3-surfaces of a kinematically possible NIF. Such data are
restricted by the super-Hamiltonian and super-momentum constraints
(which are four of Einstein's equations).

\medskip

An Einstein space-time $M^4$ (a {\it 4-geometry}) \emph{is the
equivalence class of all the completely fixed gauges (NIF) with
gauge equivalent Cauchy data for the DOs on the associated Cauchy
and simultaneity surfaces $\Sigma_{\tau}$.}

Once a solution of the hyperbolic Hamilton equations (viz. the
Einstein equations after a complete gauge fixing) has been found
corresponding to a set of Cauchy data, in each NIF we know the DOs
in that gauge ({\it the tidal effects}) and then the explicit form
of the gauge variables ({\it the inertial effects}). Moreover, {\it
the extrinsic curvature of the simultaneity surfaces $\Sigma_{\tau}$
is determined too}. Since the simultaneity surfaces are
asymptotically flat, it is possible to determine their embeddings
$z^{\mu}(\tau ,\sigma^r)$ in $M^4$. As a consequence, \emph{unlike
special relativity}, the conventions for clock synchronization and
the whole chrono-geometrical structure of $M^4$ (gravito-magnetism,
3-geodesic spatial distance on $\Sigma_{\tau}$, trajectories of
light rays in each point of $M^4$, one-way velocity of light) are
{\it dynamically determined }.

Let us remark that, if we look at Minkowski space-time as a special
solution of Einstein's equations with $R_{\bar a}(\tau ,\sigma^r) =
\Pi_{\bar a}(\tau ,\sigma^r) = 0$ (zero Riemann tensor, no tidal
effects, only inertial effects), we find \cite{20} that the
dynamically admissible 3+1 splittings (non-inertial frames) must
have the simultaneity surfaces $\Sigma_{\tau}$ {\it 3-conformally
flat}, because the conditions $R_{\bar a}(\tau ,\sigma^r) =
\Pi_{\bar a}(\tau ,\sigma^r) = 0$ imply the vanishing of the
Cotton-York tensor of $\Sigma_{\tau}$. Instead, in special
relativity, considered as an \emph{autonomous theory}, all the
non-inertial frames compatible with the M$\o$ller conditions are
admissible, so that there is much more freedom in the conventions
for clock synchronization.

\bigskip

A first application of this formalism \cite{39} has been the
determination of post-Minkowskian background-independent
gravitational waves in a completely fixed non-harmonic
3-orthogonal gauge with diagonal 3-metric. It can be shown that
the requirements $R_{\bar a}(\tau ,\sigma^r) << 1$, $\Pi_{\bar
a}(\tau ,\sigma^r) << 1$ lead to a weak field approximation based
on a Hamiltonian linearization scheme:

\noindent i) linearize the Lichnerowicz equation, determine the
conformal factor of the 3-metric and then the  lapse and shift
functions;

\noindent ii) find $E_{ADM}$ in this gauge and disregard all the
terms more than quadratic in the DOs;

\noindent iii) solve the Hamilton equations for the DOs.

In this way we get a solution of linearized Einstein's equations, in
which the configurational DOs $R_{\bar a}(\tau ,\sigma^r)$ play the
role of the {\it two polarizations} of the gravitational wave and we
can evaluate the embedding $z^{\mu}(\tau ,\sigma^r)$ of the
simultaneity surfaces of this gauge explicitly.

\bigskip

Let us conclude with some remarks about the interpretation of the
space-time 4-manifold in general relativity.
\medskip

In 1914 Einstein, during his researches for developing general
relativity, faced the problem arising from the fact that the
requirement of general covariance would involve a threat to the
physical objectivity of the points of space-time $M^4$, which in
classical field theories are usually assumed to have a {\it well
defined individuality}. This led him to formulate the \emph{Hole
Argument}. Assume that $M^4$ contains a {\it hole} ${\cal H}$, that
is an open region where all the non-gravitational fields vanish. It
is implicitly assumed that the Cauchy surface for Einstein's
equations lies outside ${\cal H}$. Let us consider an active
diffeomorphism $A$ which re-maps the points inside ${\cal H}$, but
is the identity outside ${\cal H}$. For any point $p \in {\cal H}$
we have $p \mapsto D_A\, p \in {\cal H}$. The induced active
diffeomorphism on the 4-metric tensor ${}^4g$, solution of
Einstein's equations, will map it into another solution $D^*_A\,
{}^4g$ ($D^*_A$ is a dynamical symmetry of Einstein's equations)
defined by $D^*_A\, {}^4g(D_A\, p) = {}^4g(p) \not= D^*_A\,
{}^4g(p)$. Consequently, we get two solutions of Einstein's
equations with the same Cauchy data outside ${\cal H}$ and it is not
clear how to save the identification of the mathematical points of
$M^4$.

\bigskip

Einstein avoided the problem by means of the pragmatic {\it
point-coincidence argument}: the only real world-occurrences are the
(coordinate-independent) space-time coincidences (like the
intersection of two world-lines). However, the problem was rebirth
by Stachel \cite{40} and then by Earman and Norton \cite{41}, and
this opened a rich philosophical debate that is still alive today.

We must face the following dilemma:

If we insist on the reality of space-time mathematical points
independently of the presence of any physical field (the {\it
substantivalist} point of view of philosophers), we are in trouble
with predictability.

If we say that ${}^4g$ and $D^*_A\, {}^4g$ describe the same
universe (the so-called {\it Leibniz equivalence}), we lose any
physical objectivity of the space-time points (the {\it
anti-substantivalist} point of view).

Stachel \cite{40} suggested that a physical individuation of the
point-events of $M^4$ could be made only by using {\it four
individuating fields depending on the 4-metric on $M^4$}, namely
that a tensor field on $M^4$ is needed to identify the points of
$M^4$.
\medskip

On the other hand, {\it coordinatization} is the only way to
individuate the points {\it mathematically} since, as stressed by
Hermann Weyl \cite{42}: ''There is no distinguishing objective
property by which one could tell apart one point from all others
in a homogeneous space: at this level, fixation of a point is
possible only by a {\it demonstrative act} as indicated by terms
like {\it this} and {\it there}.''

\bigskip

To clarify the situation let us remember that Bergmann and Komar
\cite{43} gave {\it a passive re-interpretation of active
diffeomorphisms as metric-dependent coordinate transformations}
$x^{\mu} \mapsto y^{\mu}(x, {}^4g(x))$ restricted to the solutions
of Einstein's equations (i.e. {\it on-shell}). It can be shown that
\emph{on-shell} ordinary passive diffeomorphisms and the
\emph{on-shell} Legendre pull-back of Hamiltonian gauge
transformations are two (overlapping) dense subsets of this set of
\emph{on-shell} metric-dependent coordinate transformations. Since
the Cauchy surface for the Hole Argument lies outside the hole
(where the active diffeomorphism is the identity), it follows that
the passive re-interpretation of the active diffeomorphism $D^*_A$
must be an {\it on-shell Hamiltonian gauge transformation}, so that
the \emph{Leibniz equivalence} reduces to gauge equivalence in the
sense of Dirac constraint theory (${}^4g$ and $D^*_A\, {}^4g$ belong
to the same gauge orbit). In our language, \emph{Leibniz
equivalence} is then reduced to a change of NIF for the same
Einstein \emph{universe}.

\bigskip

What remains to be done is to implement Stachel's suggestion
according to which  the {\it intrinsic pseudo-coordinates} of
Bergmann and Komar \cite{44} should be used as \emph{individuating
fields}. These pseudo-coordinates for $M^4$ (at least when there are
no Killing vectors) are four scalar functions $F^A[w_{\lambda}]$,
$A, \lambda = 1,..,4$, of the four eigenvalues $w_{\lambda}({}^4g,
\partial\, {}^4g)$ of the spatial part of the Weyl tensor. Since
these eigenvalues can be shown to be in general functions of the
3-metric, of its conjugate canonical momentum (namely of the
extrinsic curvature of $\Sigma_{\tau}$) and of the lapse and shift
functions, the pseudo-coordinates are well defined in phase space
and can be used as a label for the points of $M^4$.

\bigskip

The final step \cite{27} is to implement the individuation of
point-events by considering an arbitrary kinematically admissible
3+1 splitting of $M^4$ with a given time-like observer and the
associated radar 4-coordinates $\sigma^A$ (a NIF), and imposing the
following gauge fixings to the secondary super-Hamiltonian and
super-momentum constraints (the only restriction on the functions
$F^A$ is the orbit condition):

\begin{equation}
 \chi^A(\tau ,\sigma^r) = \sigma^A - F^A[w_{\lambda}] \approx 0.
\end{equation}

In this way we break general covariance completely and we determine
the gauge variables $\theta^r$ and $\pi_{\phi}$. Then the
$\tau$-constancy of these gauge fixings will produce the gauge
fixings determining the lapse and shift functions. After having
fixed the Lorentz gauge freedom of the cotetrads, we arrive at a
completely fixed gauge in which, after the transition to Dirac
brackets, we get $\sigma^A \equiv {\tilde F}^A[r_{\bar a}(\sigma ),
\pi_{\bar a}(\sigma )]$, namely the conclusion that the \emph{radar}
4-coordinates of a point in $M^4_{3+1}$, the copy of $M^4$
coordinatized with the chosen non-inertial frame, are determined
{\it off-shell} by the four DOs of that gauge: in other words {\it
the \emph{individuating fields} nothing else than are the genuine
\emph{tidal} effects of the gravitational field}. By varying the
functions $F^A$ we can make an analogous off-shell identification in
every other admissible non-inertial frame. The procedure is
consistent, because the DOs are functionals of the metric and the
extrinsic curvature on a whole 3-space $\Sigma_{\tau}$ but in fact
know the whole 3+1 splitting $M^4_{3+1}$ of $M^4$.

\bigskip

Some consequences of this identification of the point-events of
$M^4$ are:\medskip

1) \emph{The physical space-time $M^4$ and the vacuum gravitational
field are essentially the same entity}. The presence of matter
modifies the solutions of Einstein equations, i.e. $M^4$, but plays
only an indirect role in this identification (see Ref.\cite{27}. On
the other hand, matter is fundamental in establishing a (still
lacking) dynamical theory of measurement exploiting non-test
objects. Consequently, instead of the dichotomy
substantivalism/relationism, it seems that this analysis - as a case
study limited to the class of space-times dealt with - may offer a
new more articulated point of view, which can be named {\it point
structuralism} (see Ref. \cite{45}).

\medskip

2) The \emph{reduced phase space} of this model of general
relativity is the space of \emph{abstract} DOs (pure \emph{tidal}
effects without \emph{inertial} effects), which can be thought of as
four fields residing on an abstract space-time ${\tilde M}^4$
defined as the equivalence class of all the admissible, non-inertial
frames $M^4_{3+1}$ containing the associated inertial effects.

\medskip

3) Each \emph{radar} 4-coordinate system of an admissible
non-inertial frame $M^4_{3+1}$ has an associated {\it
non-commutative structure}, determined by the Dirac brackets of the
functions $ {\tilde F}^A[r_{\bar a}(\sigma ), \pi_{\bar a}(\sigma
)]$ determining the gauge, a fact that could play a role in the
quantization of the theory.

\bigskip

As a final remark, let us note that these results on the
identification of point-events are {\it model dependent}. In
spatially compact space-times without boundary, the DOs are {\it
constants of the motion} due to the frozen picture. As a
consequence, the gauge fixings $\chi^A(\tau ,\sigma^r) \approx 0$
(in particular $\chi^{\tau}$) cannot be used to rebuild the temporal
dimension: probably only the instantaneous 3-space of a 3+1
splitting can be individuated in this way.

\vfill\eject

\section{Conclusions.}

Our everyday experience of macroscopic objects and processes is
scientifically described in terms of Newtonian physics with its
separate notions of time (and simultaneity) and (Euclidean
instantaneous) 3-space. A huge amount of philosophical literature
has been devoted to the analysis of the consequences that follow
from the empirical fact that light velocity, as well as that of any
causal propagation, has finite magnitude. Our macroscopic experience
is dominated by Maxwell equations even for the fact that, from the
physical and neurophysiological point of view, all the information
that reaches our brain is of electro-magnetic origin. Therefore the
consequences of the finite magnitude of the causal propagation of
energy and information has a direct bearing on our phenomenological
experience. On the other hand, the conventional nature of the
definition of distant simultaneity that follows from the analysis of
the basic structure of causal influences in SR seems to conflict
with every possible notion of {\it 3-dimensional reality} of objects
and processes which stands at the basis of our phenomenological
experience since it entails that no observer- and frame-independent
notions of simultaneity and instantaneous 3-space be possible. Even
if - from the technical point of view - the question of the
conventionality of simultaneity can be rephrased as a gauge problem,
it lasted as source of an unending debate involving old fundamental
issues concerning the philosophy of time like that of the nature of
now-ness, becoming, reality or unreality of time, past and future,
with all possible ramifications and varieties of philosophical
distinctions.

\medskip

It should not be undervalued that relativistic thinking unifies the
\emph{physical} notions of space and time in a 4-dimensional
structure, whilst space and time maintain a substantial ontological
diversity in our phenomenological experience. While time is
experienced as "flowing", space is not. Furthermore time, even more
than space, plays a fundamental constitutive role for our "being in
the world" and for \emph{subjectivity} in general, which manifests
itself in living beings with various gradations. There is,
therefore, a deep contrast between the formal inter-subjective
unification of space and time in the scientific relativistic image,
on the one hand, and the ontological diversity of time and space
within the subjectivity of experience, on the other. This appears to
be the most important and difficult question that physics raises to
contemporary philosophy, since it reveals the core of the relation
between \emph{reality} of \emph{experience} and
\emph{reality-objectivity} of \emph{knowledge}. Dismissing this
contrast by a literal adoption of the scientific image is not as
much a painless and obvious operation as rather an implicit adoption
of a strong physicalist philosophical position that should be argued
for itself.

\medskip

This said, we have faced the question to investigate a possible
contribution of the inclusion of gravity (which, as well-known, is a
universal interaction that cannot be shielded) to the clarification
of the problem of relativistic distant simultaneity. This has been
done having in view certainly not a \emph{technical resolution} of
the above philosophical contrast, rather as the achievement of a
notion of distant simultaneity within the scientific image which be
at least \emph{compatible} with our deep experience of what
Whitehead called the "cosmic unison".

\medskip

As a matter of fact, \emph{we have shown that the inclusion of
gravity deeply changes the state of affairs about relativistic
simultaneity}.

\medskip

In brief, we have identified a class of curved pseudo-Riemannian
space-times in which the following results holds:\medskip

i) Outside the solutions of Einstein's equations (i.e.
\emph{off-shell}), these space-times admit 3+1 splittings, which can
be interpreted as \emph{kinematically possible global non-inertial
laboratories} (kinematically possible NIFs) centered on
\emph{arbitrary accelerated observers}. The viewpoint following from
this concept leads to a frame-dependent notion of
\emph{instantaneous 3-space,} which is concomitantly a \emph{clock
synchronization convention}. As in SR, all these conventions are
gauge equivalent, so that there is \emph{no Wheeler-DeWitt
interpretation} of the gauge transformations generated by the
super-Hamiltonian constraint.\medskip

ii) The \emph{off-shell} Hamiltonian separation of the\emph{ tidal
}degrees of freedom of the gravitational field from the \emph{gauge
variables} implies the interpretation of the latter as
\emph{relativistic inertial effects} which are shown in the chosen
kinematical NIF. Since in this class of space-times the Hamiltonian
is the weak ADM energy plus a combination of the first class
constraints, in every completely fixed gauge (a well defined
kinematical NIF) it follows deterministic evolution of the
\emph{tidal degrees of freedom} in mathematical time (to be replaced
by a physical clock when eventually possible) governed by the
\emph{tidal forces }and the \emph{inertial forces} of that NIF (note
that unlike the Newtonian physics, such forces are in general
functions of the tidal degrees of freedom too)\footnote{In these
globally hyperbolic space-times there is no \emph{frozen picture of
dynamics}.}.

\medskip

iii) The solution of Hamilton equations in a completely fixed gauge
with given Cauchy data for the \emph{tidal} degrees of freedom (and
matter if present) determines a solution of Einstein's equations in
a well defined 4-coordinated system, which \emph{on-shell} are
re-interpretable as coordinates adapted to a \emph{dynamically
determined} NIF (one of its leaves is the Cauchy surface on which
the Cauchy data have been assigned).

\medskip

iv) Given any solution of Einstein's equations in a given
4-coordinate system, we can determine the dynamical 3+1 splitting (a
\emph{dynamical} NIF) of Einstein's space-time, one of whose
simultaneity 3-surfaces is just the Cauchy surface of the solution.
Consequently, there is a {\it dynamical emergence of the
instantaneous 3-spaces}, leaves of the dynamical NIF, for each
solution of Einstein's equations in a given 4-coordinate system
(adapted \emph{on-shell} to the dynamical NIF). Moreover, all the
chrono-geometrical structure of Einstein's space-time ($ds^2 =
{}^4g_{\mu\nu}(x)\, dx^{\mu}\, dx^{\nu}$) is \emph{dynamically
determined}.

\medskip

v) These results and a revisitation of the Hole Argument imply that
{\it space-time and vacuum gravitational field are two faces of the
same reality}, and we get a \emph{new kind of structuralism } (with
elements of both the \emph{substantivalist} and \emph{relationist}
points of view) implying a 4-dimensional holism (see Ref.\cite{27})
resulting from a foliation with 3-dimensional instantaneous 3-spaces
\footnote{In spatially compact space-times without boundary, where
there is a frozen picture of dynamics and only a local
time-evolution according to the Wheeler-DeWitt interpretation, only
3-space, but not the time direction, can be determined from the
gravitational field.}.

\bigskip

In conclusion what in Newton's theory was an absolute Euclidean
instantaneous 3-space reappears in GR as a \emph{dynamicadlly
emergent} Riemannian time-varying instantaneous 3-space, which is a
simultaneity leaf of a \emph{dynamical} NIF \emph{uniquely
associated to a solution of Einstein's equations in 4-coordinates
adapted to the NIF itself.} The NIF is centered on a time-like (in
general accelerated) observer, whose world-line can be made to
coincide with the Fokker-Pryce center of inertia by means of a
suitable gauge fixing to the rest-frame conditions. In the
post-Newtonian approximation around the Earth we describe the
situation in a \emph{quasi-inertial frame} with harmonic
4-coordinates, as those considered in the IAU conventions for the
geocentric celestial reference frame \cite{54}. However, in this
case too the IAU frame is not the dynamical NIF associated with the
post-Newtonian solution of the Einstein's equations: it has to be
determined through the inverse problem starting from the
post-Newtonian extrinsic curvature 3-tensor.

\medskip

Admittedly, all the physical implications of this viewpoint must
still be worked out (for instance the determination of non-inertial
frames in which the Riemannian distance from the Earth to a galaxy
equals the galaxy luminosity distance and the implications for dark
matter and dark energy of the dynamical instantaneous 3-spaces).

\medskip

Let us remark that in SR (and in GR too before identifying the
preferred dynamical convention of clock synchronization), an ideal
observer has the following freedom in the description of the
phenomena around him:

a) the arbitrary choice of the clock synchronization convention,
i.e., of the instantaneous 3-spaces;

b) the choice of the 4-coordinate system.

After these choices, the observer has a description of the other
world-lines and/or world-tubes simulating the phenomena with
Hamiltonian evolution in the chosen time parameter. All these
descriptions have been shown to be gauge-equivalent in the previous
sections. Every other ideal observer has the same type of freedom in
the description of the phenomena.

\medskip

Each solution of Einstein's equations, i.e. each Einstein
\emph{universe}, in our class of models, is an equivalence class of
well-defined \emph{dynamical} NIFs (the \emph{epistemic part} of the
metric field describing the \emph{generalized relativistic inertial
effects}) with their \emph{dynamical clock synchronization
conventions}, their \emph{dynamical instantaneous 3-spaces }and
their \emph{dynamical individuation of point-events}\footnote{Maybe
even realized by means of 4-coordinates not adapted to the NIF; we
simplified the exposition by formulating the NIFs with adapted
\emph{radar} 4-coordinates}. The NIFs selected by one solution are
\emph{different} from the NIFs selected by a \emph{different
solution}. Let us stress, however, that given a solution, the set of
the associated NIFs is a \emph{substantially smaller} set than that
of the a-priori \emph{kinematically} possible NIFs both in GR and in
SR, since the only restrictions at the kinematical level are given
by the M$\o$ller conditions\footnote{If Minkowski space-time without
matter is considered as a special solution of Einstein's equations,
its dynamical NIFs have the simultaneity leaves 3-conformally flat
\cite{20,26}.}. Given an Einstein \emph{universe}, all the
associated NIFs in the equivalence class are connected by
\emph{on-shell Hamiltonian gauge transformations} (containing
adapted passive diffeomorphisms) so that they know - as it were -
the Cauchy data of the solution. Note moreover that they also
contain the freedom of changing the time-like observer's origin of
the 3-coordinates on the instantaneous 3-spaces, and the freedom of
making an arbitrary (tensorial) passive diffeomorphism leading to
non-adapted 4-coordinates.

\medskip

From the point of view of the 4-dimensional picture with the freedom
of passive diffeomorphisms one could be led to adopt the misleading
notion of "block universe" even in GR. However, one should not
forget that this "block universe" is the equivalence class of
\emph{dynamical} NIFs (stratification or 3+1 splittings with
dynamical generated instantaneous 3-spaces). Once the epistemic
framework of a NIF is chosen, a \emph{well-defined B-series is
established}, since the notion of "earlier than", "later than" and
"simultaneous with" is globally defined in terms of the mathematical
time of the globally hyperbolic space-time (to be then replaced with
a physical clock monotonically increasing in the mathematical time,
at least for a finite interval). Unlike in SR, where each observer
has their own B-series, we have disclosed the relevant fact that, at
least for a specific class of models, GR is characterized by a
\emph{universal B-series}. As a matter of fact, a Hamiltonian gauge
transformation changes the NIF, the Hamiltonian, the matter
distribution, the Cauchy surface and the form in which the Cauchy
data are given on it, in such a way that the B-series relations
between \emph{any pair of events} is left invariant. The same
happens with the freedom of changing the time-like observer that
defines the origin of the coordinates on the instantaneous 3-spaces.
\medskip

Nothing are we willing to say about the relevance of A-series
determinations within the scientific image, as already stressed in
the Introduction. We maintain that any tensed determination is
wholly foreign to any kind of physical description of the world.
Clearly, unlike the case of SR, we here have Hamiltonian evolution
of the 3-space too, determined by the ADM energy (which depends on
the \emph{ontic} tidal effects and on the matter). This evolution,
however - as to its temporal characterization - is not substantially
different from the Newtonian evolution of matter in absolute time.

\medskip

It remains to be clarified - as anticipated in the Introduction -
the import (if any) of our technical results on the dichotomy
\emph{endurantism/perdurantism}. Having already stated our
definition of \emph{wholly presentness} of a \emph{physical object}
in the Introduction, we must now only ascertain which of the
relevant physical features of a general-relativistic space-time with
matter we have enumerated so far, tend to support a reasonable
notion of \emph{endurantism} or \emph{perdurantism} at the level of
the scientific image.

It seems clear that all the attributes which are necessary to define
a spatially-extended \emph{physical object} belonging to a
dynamically generated 3-space $\Sigma_{\tau}$ at a certain time
$\tau$, can be obtained by the chrono-geometrical features intrinsic
to $\Sigma_{\tau}$ and matter distribution on it. Note - remarkably
- that among such attributes there is the 3-geometry of
$\Sigma_{\tau}$. We can conclude, therefore, that, in this limited
sense, our results support an \emph{endurantist} view of physical
objects. It is interesting to note, on the other hand, that our
analysis does not support a likewise simple \emph{endurantist} view
of the space-time structure itself. We have established that the
\emph{reality} of the vacuum space-time of GR is ontologically
equivalent to the \emph{reality }of the autonomous degrees of
freedom of the gravitational field as described by the DOs (viz.,
the \emph{ontic part} of the metric field). At this point we should
look at space-time itself as at something sharing the attributes of
a peculiar \emph{physical object}. We could ask accordingly whether
and to what extent the mathematical structure of the DOss allows an
\emph{endurantist} interpretation of such a peculiar object. The
answer is simple: as already said the DOs, though being local fields
indexed by the \emph{radar} coordinates $\sigma^A$, when considered
in relation to the 4-metric field $g$, are highly non-local
functionals of the 3-metric field and of the extrinsic curvature
3-tensor on the whole 3-space $\Sigma_{\tau}$\footnote{Let us recall
that this 3-tensor describes the \emph{embedding} of the
instantaneous 3-space in the Einstein's 4-manifold through its
dependence upon the \emph{$\tau$-derivative} of the 3-metric and the
\emph{lapse} and \emph{shift} functions.}. Due to the extrinsic
curvature, the structure of the DOs involves therefore an
infinitesimal $\tau$-continuum of 3-spaces around $\Sigma_{\tau}$.
The individuation procedure involves moreover a \emph{temporal
gauge} (A=0 in Equation (4.3)). In conclusion, the \emph{physical
individuation} of the space-time \emph{point-events}, defined by
Equation (4.3), \emph{cannot} be considered as an attribute
depending upon information wholly contained in the 3-space
considered at time $\tau$. This conclusion, however, should not be
viewed as an unexpected and unsatisfactory result, given the double
role of the metric field in GR.

\bigskip

Let us close by stressing a general fundamental feature of GR.
Though Einstein's partial differential equations are defined in a
4-dimensional framework, this framework must be considered as an
unfolding of 3-dimensional sub-structures because of the nature of
the Cauchy problem. Consequently, the \emph{models} of GR are
subdivided into two disjoint classes:

a) the 4-dimensional ones (with spatially-compact space-times),
having the \emph{problem of time}, the \emph{frozen picture}, and a
lacking \emph{physical individuation} of point-events;

b) the asymptotically-flat space-times, with their 3+1 splitting and
\emph{dynamical emergence} of \emph{achronal} 3-spaces, a \emph{non
trivial temporal evolution} and a \emph{physical individuation} of
point-events.

\vfill\eject

\end{document}